\documentclass{pasj00}
\usepackage{multirow}
\usepackage{booktabs}
\usepackage{threeparttable}
\usepackage{natbib}
\usepackage{pdflscape}
\usepackage{graphicx}

\newcommand{\mjybm}{\mbox{mJy~beam${}^{-1}$}}
\newcommand{\jybm}{\mbox{Jy~beam${}^{-1}$}}

\begin{document}
\SetRunningHead{K. Niinuma et al.}{KaVA observation of bright AGN jets}

\title{VLBI observations of bright AGN jets with KVN and VERA Array (KaVA): Evaluation of Imaging Capability}

 \author{%
   Kotaro \textsc{Niinuma}\altaffilmark{1}, 
   Sang-Sung \textsc{Lee}\altaffilmark{2}, 
   Motoki \textsc{Kino}\altaffilmark{2, 3}, 
   Bong Won \textsc{Sohn}\altaffilmark{2, 4}, 
   Kazunori \textsc{Akiyama}\altaffilmark{5,6}, 
   Guang-Yao \textsc{Zhao}\altaffilmark{2}, 
   Satoko \textsc{Sawada-Satoh}\altaffilmark{7}, 
   Sascha \textsc{Trippe}\altaffilmark{8}, 
   Kazuhiro \textsc{Hada}\altaffilmark{6, 9},    
   Taehyun \textsc{Jung}\altaffilmark{2}, 
   Yoshiaki \textsc{Hagiwara}\altaffilmark{6, 10}, 
   Richard \textsc{Dodson}\altaffilmark{2}, 
   Shoko \textsc{Koyama}\altaffilmark{3, 5, 6,11}, 
   Mareki \textsc{Honma}\altaffilmark{6, 10}, 
   Hiroshi \textsc{Nagai}\altaffilmark{12}, 
   Aeree \textsc{Chung}\altaffilmark{13}
   Akihiro \textsc{Doi}\altaffilmark{3}
   Kenta \textsc{Fujisawa}\altaffilmark{14}
   Myoung-Hee \textsc{Han}\altaffilmark{2},    
   Joeng-Sook \textsc{Kim}\altaffilmark{6}, 
   Jeewon \textsc{Lee}\altaffilmark{2, 15}, 
   Jeong Ae \textsc{Lee}\altaffilmark{2},
   Atsushi \textsc{Miyazaki}\altaffilmark{2},    
   Tomoaki \textsc{Oyama}\altaffilmark{6},    
   Kazuo \textsc{Sorai}\altaffilmark{16}, 
   Kiyoaki \textsc{Wajima}\altaffilmark{2}, 
   Jaehan \textsc{Bae}\altaffilmark{2, 17}, 
   Do-Young \textsc{Byun}\altaffilmark{2}, 
   Se-Hyung \textsc{Cho}\altaffilmark{2}, 
   Yoon Kyung \textsc{Choi}\altaffilmark{2}, 
   Hyunsoo \textsc{Chung}\altaffilmark{2}, 
   Moon-Hee \textsc{Chung}\altaffilmark{2}, 
   Seog-Tae \textsc{Han}\altaffilmark{2}, 
   Tomoya \textsc{Hirota}\altaffilmark{6}, 
   Jung-Wook \textsc{Hwang}\altaffilmark{2}, 
   Do-Heung \textsc{Je}\altaffilmark{2}, 
   Takaaki \textsc{Jike}\altaffilmark{7}, 
   Dong-Kyu \textsc{Jung}\altaffilmark{2}, 
   Jin-Seung \textsc{Jung}\altaffilmark{2}, 
   Ji-Hyun \textsc{Kang}\altaffilmark{2}, 
   Jiman \textsc{Kang}\altaffilmark{2}, 
   Yong-Woo \textsc{Kang}\altaffilmark{2}, 
   Yukitoshi \textsc{Kan-ya}\altaffilmark{6}, 
   Masahiro \textsc{Kanaguchi}\altaffilmark{6}, 
   Noriyuki \textsc{Kawaguchi}\altaffilmark{7, 18}, 
   Bong Gyu \textsc{Kim}\altaffilmark{2}, 
   Hyo Ryoung \textsc{Kim}\altaffilmark{2}, 
   Hyun-Goo \textsc{Kim}\altaffilmark{2}, 
   Jaeheon \textsc{Kim}\altaffilmark{2}, 
   Jongsoo \textsc{Kim}\altaffilmark{2}, 
   Kee-Tae \textsc{Kim}\altaffilmark{2}, 
   Mikyoung \textsc{Kim}\altaffilmark{2}, 
   Hideyuki \textsc{Kobayashi}\altaffilmark{6}, 
   Yusuke \textsc{Kono}\altaffilmark{6}, 
   Tomoharu \textsc{Kurayama}\altaffilmark{19}, 
   Changhoon \textsc{Lee}\altaffilmark{2}, 
   Jung-Won \textsc{Lee}\altaffilmark{2}, 
   Sang Hyun \textsc{Lee}\altaffilmark{2}, 
   Young Chol \textsc{Minh}\altaffilmark{2}, 
   Naoko \textsc{Matsumoto}\altaffilmark{6}, 
   Akiharu \textsc{Nakagawa}\altaffilmark{20}, 
   Chung Sik \textsc{Oh}\altaffilmark{2}, 
   Se-Jin \textsc{Oh}\altaffilmark{2}, 
   Sun-Youp \textsc{Park}\altaffilmark{2}, 
   Duk-Gyoo \textsc{Roh}\altaffilmark{2}, 
   Tetsuo \textsc{Sasao}\altaffilmark{2,6,21}, 
   Katsunori M. \textsc{Shibata}\altaffilmark{6, 10}, 
   Min-Gyu \textsc{Song}\altaffilmark{2}, 
   Yoshiaki \textsc{Tamura}\altaffilmark{7}, 
   Seog-Oh \textsc{Wi}\altaffilmark{2}, 
   Jae-Hwan \textsc{Yeom}\altaffilmark{2}, and
   Young Joo \textsc{Yun}\altaffilmark{2}
}

 \altaffiltext{1}{Graduate School of Science and Engineering, Yamaguchi University, Yoshida 1677-1, Yamaguchi, Yamaguchi 753-8512, Japan}
 \email{niinuma@yamaguchi-u.ac.jp}
 \altaffiltext{2}{Korea Astronomy and Space Science Institute, Daedeokdae-ro 776, Yuseong-gu, Daejeon 305-348, Republic of Korea}
 \altaffiltext{3}{The Institute of Space and Astronautical Science, Japan Aerospace Exploration Agency, 3-1-1 Yoshinodai, Chuou-ku, Sagamihara, Kanagawa 229-8510, Japan}
 \altaffiltext{4}{Department of Astronomy \& Space Science, University of Science \& Technology, 217 Gajeong-ro, Daejeon, Republic of Korea}
 \altaffiltext{5}{Department of Astronomy, Graduate School of Science, The University of Tokyo, 7-3-1 Hongo, Bunkyo-ku, Tokyo 113-0033, Japan}
 \altaffiltext{6}{Mizusawa VLBI Observatory, National Astoronomical Observatory of Japan, 2-21-1 Osawa, Mitaka, Tokyo 181-8588, Japan}
 \altaffiltext{7}{Mizusawa VLBI Observatory, National Astronomical Observatory of Japan, 2-12 Hoshigaoka-cho, Mizusawa-ku, Oshu, Iwate 023-0861, Japan}
 \altaffiltext{8}{Department of Physics and Astronomy, Seoul National University, Seoul 151-742, Republic of Korea}
 \altaffiltext{9}{INAF - Istituto di Radioastronomia, via Gobetti 101, I-40129 Bologna, Italy}
 \altaffiltext{10}{Department of Astronomical Science, The Graduate University of Advanced Studies (SOKENDAI), 2-21-1 Osawa, Mitaka, Tokyo 181-8588, Japan}
 \altaffiltext{11}{Max-Planck-Institut f\"{u}r Radioastronomie, Auf dem H\"{u}gel 69, Bonn, 53121, Germany}
 \altaffiltext{12}{Chile Observatory, National Astoronomical Observatory of Japan, 2-21-1 Osawa, Mitaka, Tokyo 181-8588, Japan}
 \altaffiltext{13}{Department of Astronomy, Yonsei University, 134 Shinchondong, Seodaemungu, Seoul 120-749, Republic of  Korea}
 \altaffiltext{14}{Research Institute for Time Studies, Yamaguchi University, 1677-1 Yoshida, Yamaguchi, 753-8511, Japan}
 \altaffiltext{15}{Department of Astronomy an Space Science, Kyung Hee University, Seocheon-Dong, Giheung-Gu, Yongin, Gyeonggi-Do 446-701, Republic of Korea}
 \altaffiltext{16}{Department of Cosmosciences, Graduate School of Science, Hokkaido University, Kita 10, Nishi 8, Kita-ku, Sapporo 060-0810}
 \altaffiltext{17}{Department of Astronomy, University of Michigan, 500 Church St., Ann Arbor, MI 48105, USA}
 \altaffiltext{18}{Shanghai Astronomical Observatory, Chinese Academy of Sciences 80 Nandan Road, Xuhui District, Shanghai 200030, China}
 \altaffiltext{19}{Center for Fundamental Education, Teikyo University of Science, 2525 Yatsusawa, Uenohara, Yamanashi, 409-0193, Japan}
 \altaffiltext{20}{Graduate School of Science and Engineering, Kagoshima University, 1-21-35 Korimoto, Kagoshima-shi, Kagoshima 890-0065, Japan}
 \altaffiltext{21}{Yaeyama Star Club, Ookawa, Ishigaki, Okinawa 904-0022, Japan}
 

\maketitle

\begin{abstract}
The Korean very-long-baseline interferometry (VLBI) network (KVN) and VLBI Exploration of Radio Astrometry (VERA) Array (KaVA) is the first international VLBI array dedicated to high-frequency (23 and 43 GHz bands) observations in East Asia. Here, we report the first imaging observations of three bright active galactic nuclei (AGNs) known for their complex morphologies: 4C 39.25, 3C 273, and M 87. This is one of the initial result of KaVA early science. Our KaVA images reveal extended outflows with complex substructure such as knots and limb brightening, in agreement with previous Very Long Baseline Array (VLBA) observations. Angular resolutions are better than 1.4 and 0.8 milliarcsecond at 23 GHz and 43 GHz, respectively. KaVA achieves a high dynamic range of $\sim1000$, more than three times the value achieved by VERA. We conclude that KaVA is a powerful array with a great potential for the study of AGN outflows, at least comparable to the best existing radio interferometric arrays.
\end{abstract}


\section{Introduction}
Active galactic nuclei (AGNs) are luminous emitters of radiation powered by accretion onto supermassive black holes with masses in the range from $10^6$ to $10^{10}~M_{\solar}$. A fraction of AGNs shows extended jets as large as several megaparsecs with complex structure on all spatial scales (down to subparsec). In general, AGN jets are optically thin emitters of synchrotron radiation, making them prominent at radio frequencies. Their properties imply that AGN jets are best studied through very-long-baseline interferometry (VLBI), which provides very high angular resolutions [less than one milliarcsecond (mas)] at radio frequencies. 
Recent VLBI studies provide many insights into the physical properties of AGN jets, e.g., the time variation and magnetic field in parsec-scale jets (Lister et al. 2013; O'Sullivan \& Gabuzda 2009). In addition, as over 1000 $\gamma$-ray AGNs have already been detected since the launch of the \textit{Fermi Gamma-Ray Space Telescope} in 2008 \citep{ackermann11}, numerous studies to investigate the connection between high-energy $\gamma$-ray and radio in parsec-scale jets have been attempted with various VLBI arrays (Lister et al. 2011, Nagai et al. 2010, Orienti et al. 2013). To understand the nature of AGN jets, successive and systematic studies such as these are crucial.

The capabilities of a VLBI instrument are summarised by the angular resolution and the image dynamic range (and image sensitivity). 
And it is the improvement of the image dynamic range, which is the most essential to study AGN jets. 
Improving the image dynamic range requires better Fourier coverage of the various spatial frequencies that sample the uv distribution of the target source during an observation. That is, it is expected that VLBI observations made from a number of radio telescopes would improve the image dynamic range, leading to high-quality VLBI images of target sources \citep{thompson01}.

Recently, a new VLBI facility consisting of the Korean VLBI network (KVN) and the VLBI Exploration of Radio Astrometry (VERA) has been constructed in East Asia region. 
KVN is the first VLBI array dedicated to the mm-wavelength radio observations in East Asia, and is operated by the Korea Astronomy and Space Science Institute (KASI; Lee et al. 2011; Lee et al. 2014). KVN consists of three 21-m-diameter radio telescopes: one in Seoul, one in Ulsan, and one on Jeju Island, Korea. In each, four different frequency band receivers are installed (22, 43, 86, and 129 GHz; Han et al. 2013). The baseline lengths range from 305 to 476 km. 
VERA, on the other hand, is a Japanese VLBI array dedicated to high-precision VLBI astrometry and is operated by the National Astronomical Observatory of Japan (NAOJ; Kobayashi et al. 2003). The VERA array consists of four 20-m-diameter radio telescopes with one each located at Mizusawa, Iriki, Ogasawara, and Ishigaki-jima with the baselines for these telescopes ranging from 1019 to 2270 km. Dual-beam system for efficient phase-reference VLBI observation, and 22-/43-GHz receiver system are installed.

The capabilities of KVN and VERA have already been verified as a VLBI facility \citep{lee14, honma03}, and these individual arrays are currently being used to make scientific observations. Individual VLBI arrays have already started ongoing programs to monitor AGNs. KVN initiated the program called interferometric Monitoring of Gamma-ray Bright AGNs (iMOGABA; Lee et al. 2013), and VERA started the program called Gamma-ray Emitting Notable AGN Monitoring by Japanese VLBI (\textit{GENJI}; Nagai et al. 2013).
Although these observations  have led to some progress in understanding the relationship between the radio core light curve and the $\gamma$-ray light curve, it is difficult to discuss the detailed jet structure because, being collected by only three or four stations, the image dynamic range is insufficient. 
Therefore the combined array called the KVN and VERA array (KaVA) presented herein, which consists of seven radio telescopes and a number of short baselines, is expected to provide well-filled \textit{uv}-coverage and to be sufficiently powerful to reveal the complex extended structures of AGNs. 
Although early science observation with KaVA for the 44 GHz Class I methanol maser in a massive star-forming region has already reported \citep{matsumoto14}, in order to make further observations for AGNs, the evaluation of its imaging capability is essential.

In this paper, we present the initial result of observations made with the KVN and VERA array (KaVA) of bright AGN jets. In Sections 2 and 3, we briefly introduce KaVA and discuss the individual sources selected for test observations. The details of KaVA observations and data reduction are described in Section 4, and we report those results in Section 5. In Section 6, we evaluated an imaging capability of KaVA by comparing the results of KaVA observations with those of VERA observations.
Throughout this paper, we use the cosmological parameters, $H_0=70.2~\mathrm{km~s^{-1}~Mpc^{-1}}$, $\Omega_m=0.27$, and $\Omega_\Lambda=0.73$ \citep{komatsu11}. At a distance of $z=0.01$, an angular resolution of 1 mas corresponds to a linear scale of 0.2 pc.
\section{KVN and VERA Array: KaVA}

\begin{figure}[htbp]
\begin{center}
\includegraphics[width=0.95\linewidth]{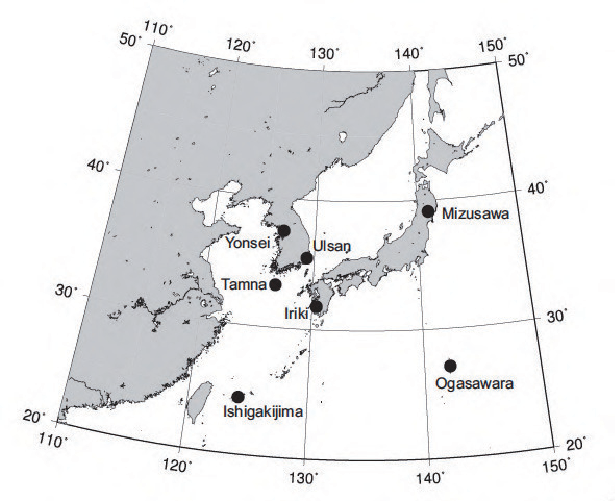} 
\end{center}
\caption{KaVA array configuration (reprinted from Figure 1 of KaVA status report).\label{fig:array}}
\end{figure}

As mentioned in the introduction, KaVA is a VLBI  array that combines KVN and VERA, and its baseline lengths range from 305 to 2270 km (Figure \ref{fig:array}, and Sawada-Satoh 2013). The observing frequencies of 23 and 43 GHz bands are available. 
Table \ref{tab:tbl0} summarizes the specifications of each array. As shown in this table, KVN has relatively short baselines ($<500$ km), and so it can detect emission from extended structures but cannot see the detailed structure on the scale of a few mas because it has insufficient angular resolution. VERA, on the other hand, has only long baselines ($>1000$ km), and so it has high angular resolution but is less sensitive to extended emission. Therefore, because each array compensates for the weakness of the other, they are complementary, which makes KaVA a very promising VLBI facility. By improving the {\it uv}-coverage, KaVA is expected to offer significant advantages for the observing extended radio sources such as AGN jets, as shown in Section \ref{obs}.

\begin{table}[htbp]
  \caption{Summary of typical array specifications of each VLBI facility.\label{tab:tbl0}}%
  \begin{center}
    \scalebox{0.9}{\begin{tabular}{lccc}\toprule\toprule
          & \multicolumn{3}{c}{Array} \\\cmidrule{2-4}
          & KVN   & VERA  & KaVA \\\midrule
    Baseline [km] & \multicolumn{1}{c}{\multirow{2}[0]{*}{305/476}} & \multicolumn{1}{c}{\multirow{2}[0]{*}{1019/2270}} & \multicolumn{1}{c}{\multirow{2}[0]{*}{305/2270}} \\
    (Shortest/Longest) & \multicolumn{1}{c}{} & \multicolumn{1}{c}{} & \multicolumn{1}{c}{} \\\addlinespace[0.05in]
    Fringe spacing [mas] & \multicolumn{1}{c}{\multirow{2}[0]{*}{5.6/3.0}} & \multicolumn{1}{c}{\multirow{2}[0]{*}{1.2/0.6}} & \multicolumn{1}{c}{\multirow{2}[0]{*}{1.2/0.6}} \\
    (23/43 GHz) & \multicolumn{1}{c}{} & \multicolumn{1}{c}{} & \multicolumn{1}{c}{} \\\addlinespace[0.05in]
    No. of Antennas & 3     & 4     & 7 \\\addlinespace[0.05in]
    No. of Baselines & 3     & 6     & 21 \\\bottomrule
    \end{tabular}}%
    \end{center}
\end{table}%

Recently, KaVA started common-use observations in both the 23 and 43 GHz bands with 1 Gbps recording (256 MHz bandwidth) in shared-risk mode. The correlation of all these 1 Gbps data recorded by further KaVA observations will be performed at the Korea-Japan Joint VLBI Correlator (KJJVC) installed at the Korea-Japan Correlation Center (KJCC) located in KASI \footnote{http://kvn.kasi.re.kr/status\_report/correlator\_status.html}.

%
\section{Source selection}\label{source}
The K4/VSOP terminal (128 Mbps recording, which equates to two-bit quantization and 32~MHz bandwidth) is os temporally being used to ensure a stable observation and correlation procedure in the commissioning phase of KaVA observation. 
Based on this recording rate, the typical baseline sensitivity ($1~\sigma$) of VERA and KaVA is expected to be 30 and 17 mJy at 23 GHz, and 63 and 26 mJy at 43 GHz. These sensitivities are derived by using the typical parameters listed in Tables 9 and 10 of the KaVA status report \footnote{http://veraserver.mtk.nao.ac.jp/restricted/KaVA2013/status\_KaVA2013.pdf} but with a bandwidth of 32~MHz. Therefore to verify the VLBI-imaging capability of KaVA, we must observe sources with significant flux.

To achieve our purpose, we selected three bright AGNs: 4C~39.25, 3C~273, and M~87. These three sources are considered to be suitable targets for verifying the imaging capability of the new VLBI array because (1) they have various characteristic structures (simple core-jet structure and extended jet) and (2) they are well studied by VLBI, so many VLBI images exist against which we can compare our results.


\begin{figure*}[htpb]

\begin{center}
\begin{minipage}{0.32\hsize}
\begin{center}
\includegraphics[width=\linewidth]{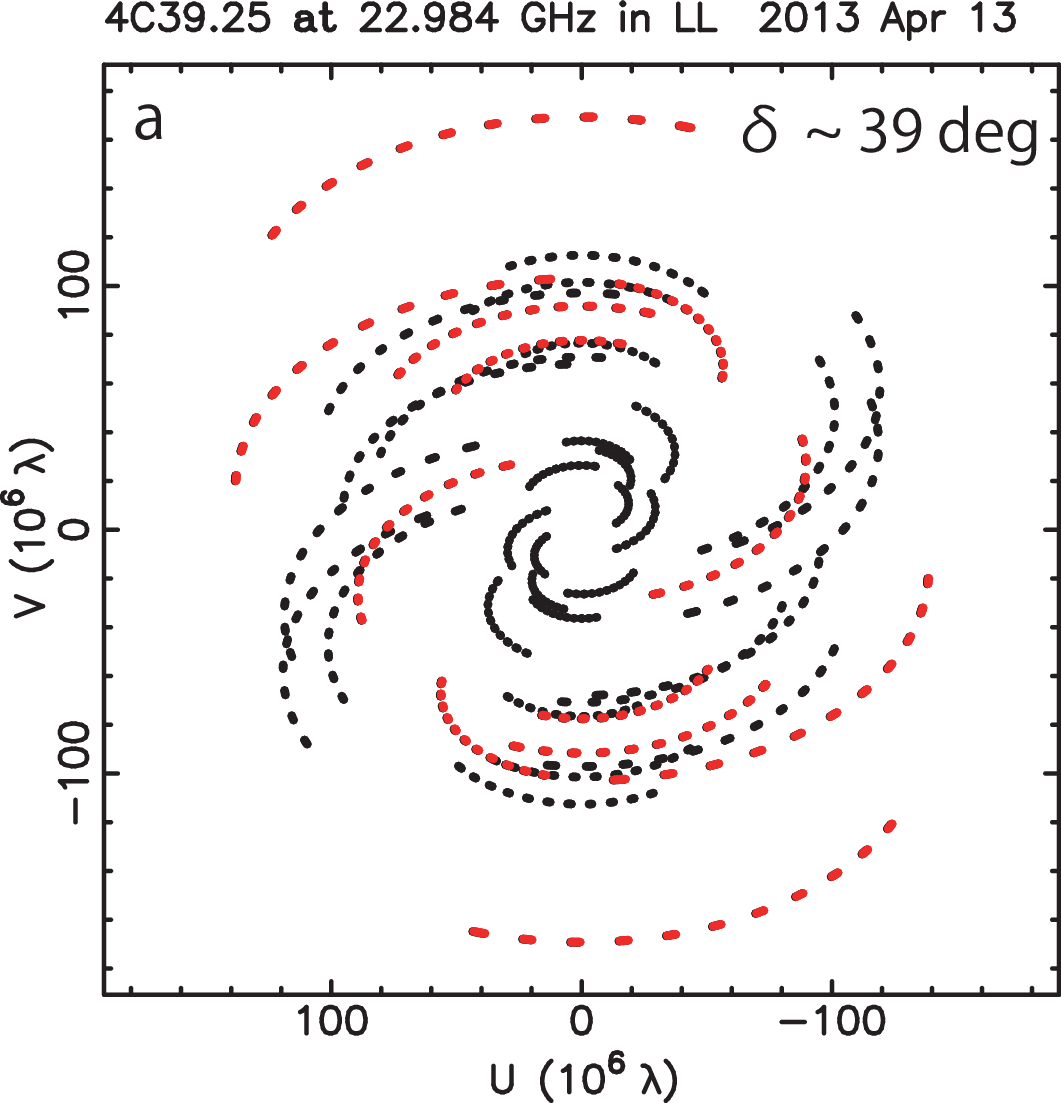}
\end{center}
\end{minipage}
\begin{minipage}{0.32\hsize}
\begin{center}
\includegraphics[width=\linewidth]{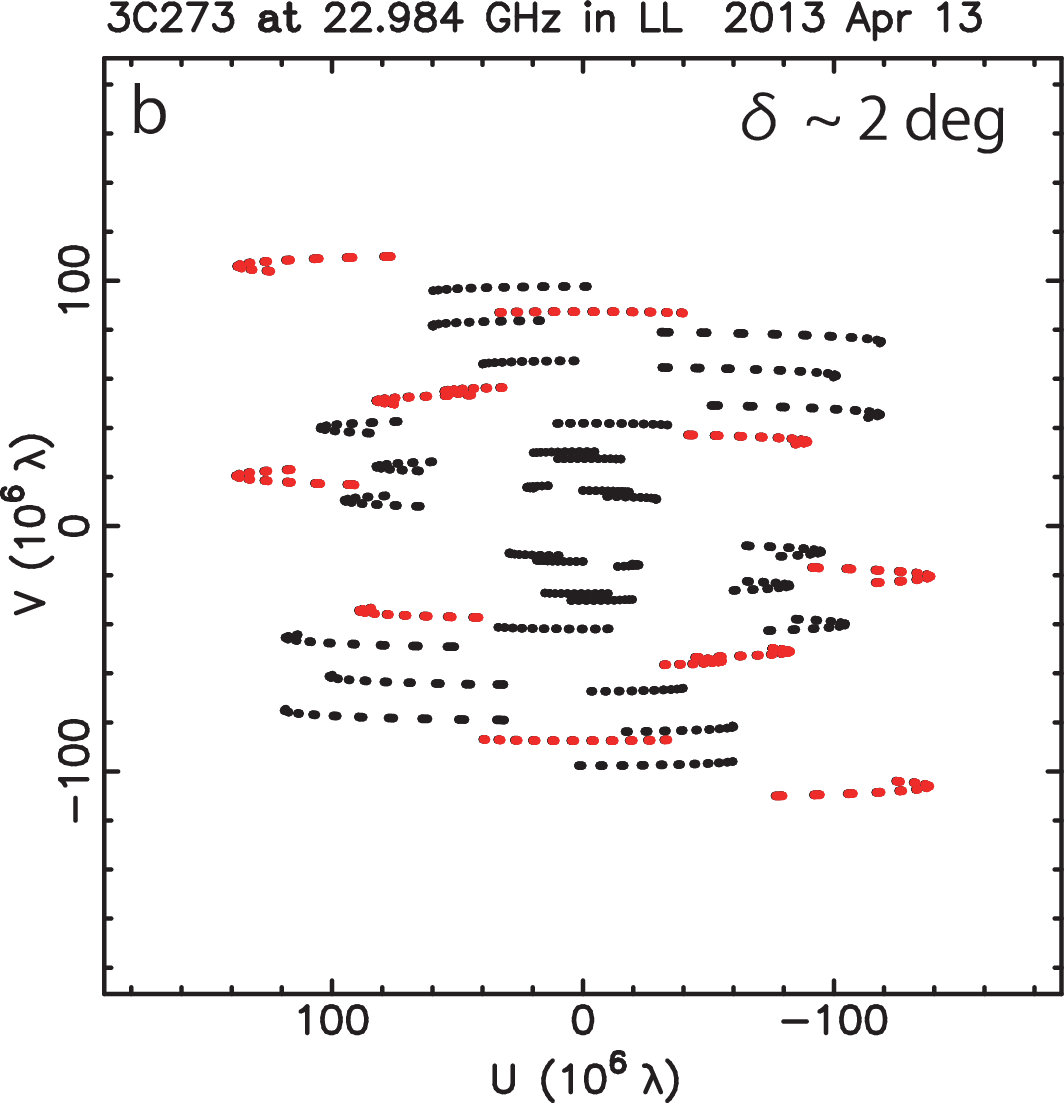}
\end{center}
\end{minipage}
\begin{minipage}{0.32\hsize}
\begin{center}
\includegraphics[width=\linewidth]{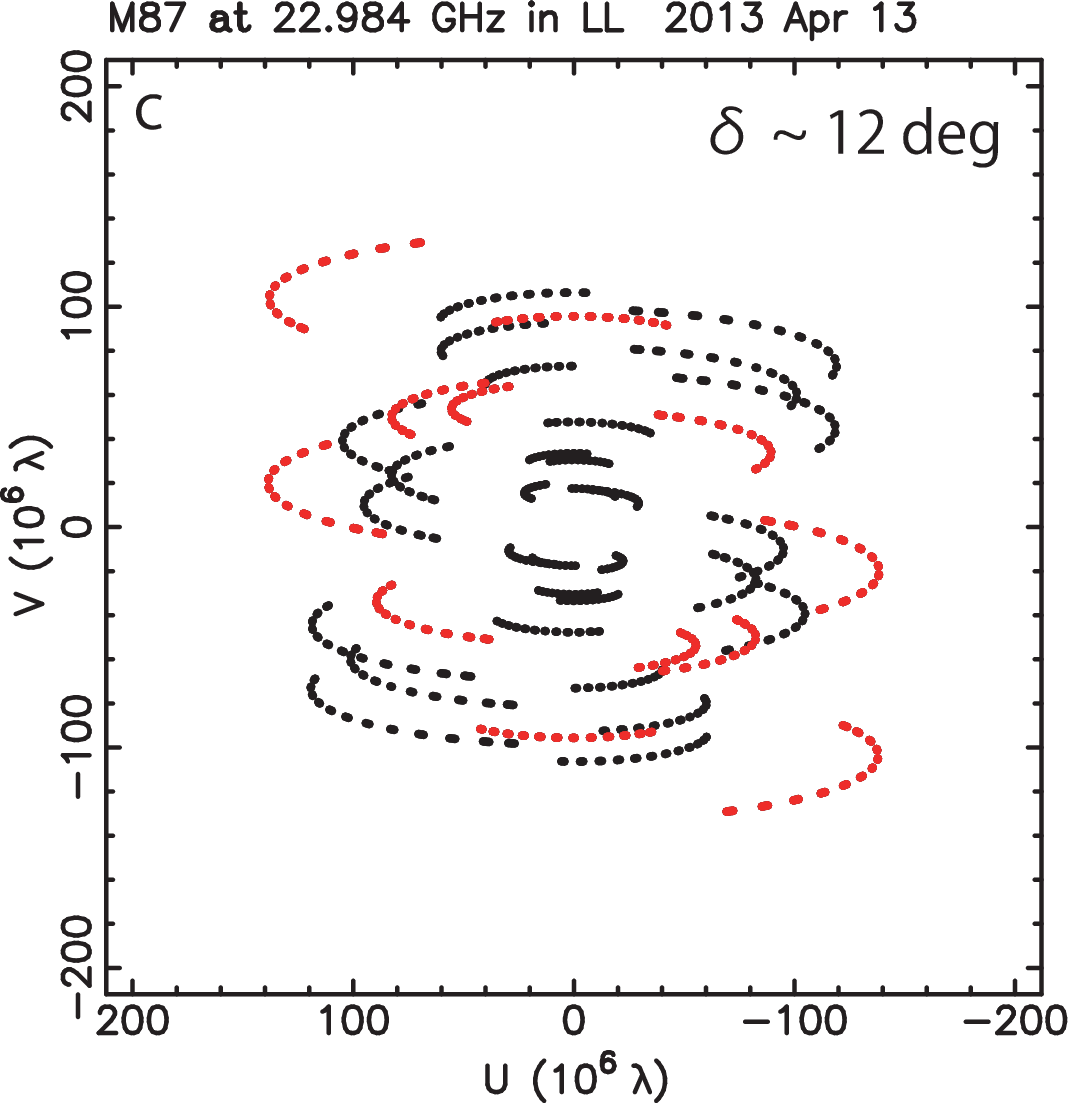}
\end{center}
\end{minipage}

\vspace{0.1in}

\begin{minipage}{0.32\hsize}
\begin{center}
\includegraphics[width=\linewidth]{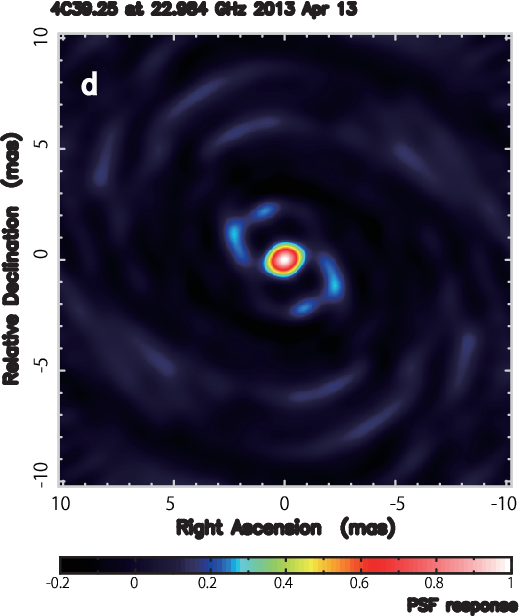}
\end{center}
\end{minipage}
\begin{minipage}{0.32\hsize}
\begin{center}
\includegraphics[width=\linewidth]{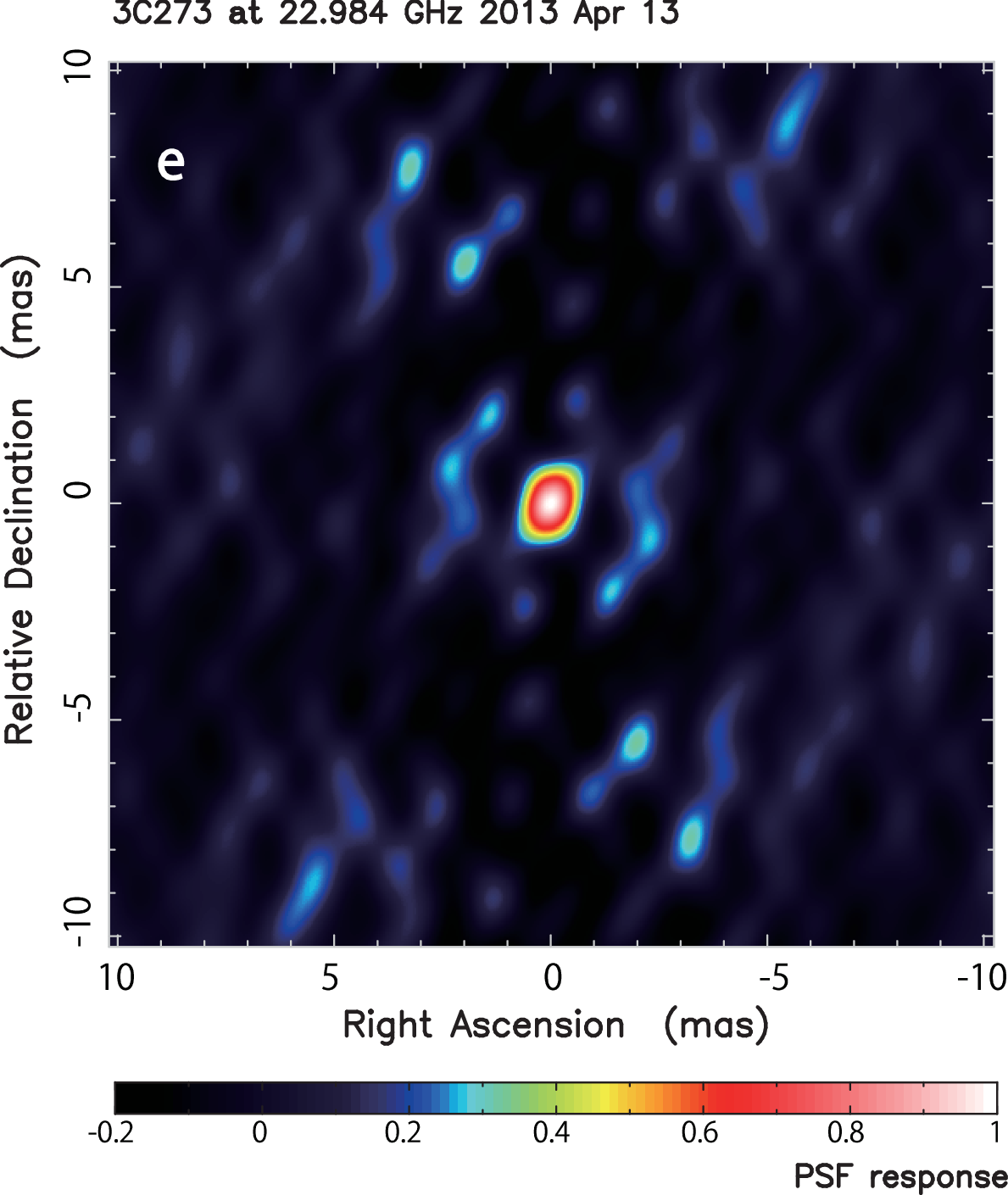}
\end{center}
\end{minipage}
\begin{minipage}{0.32\hsize}
\begin{center}
\includegraphics[width=\linewidth]{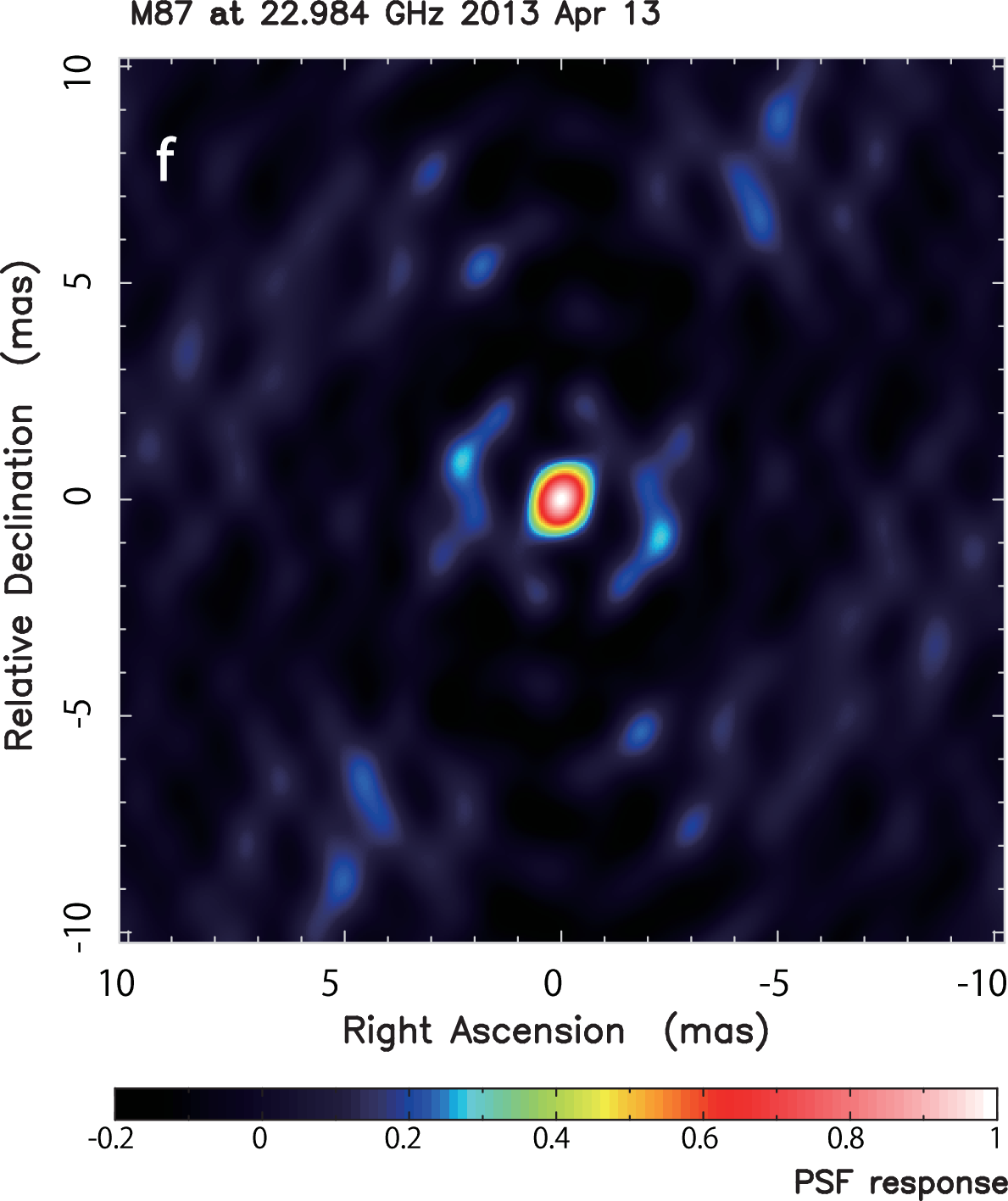}
\end{center}
\end{minipage}

\vspace{0.1in}

\begin{minipage}{0.32\hsize}
\begin{center}
\includegraphics[width=\linewidth]{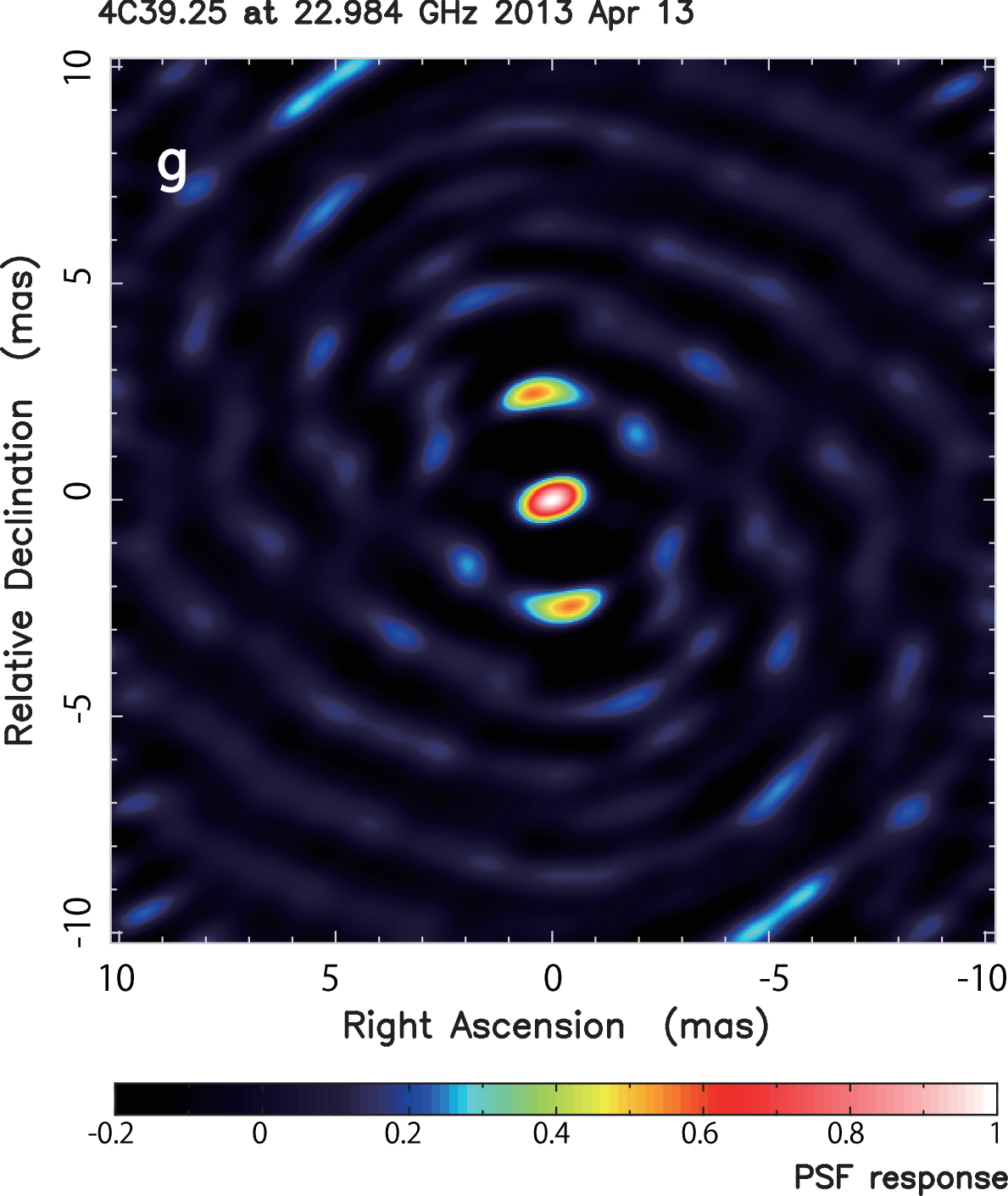}
\end{center}
\end{minipage}
\begin{minipage}{0.32\hsize}
\begin{center}
\includegraphics[width=\linewidth]{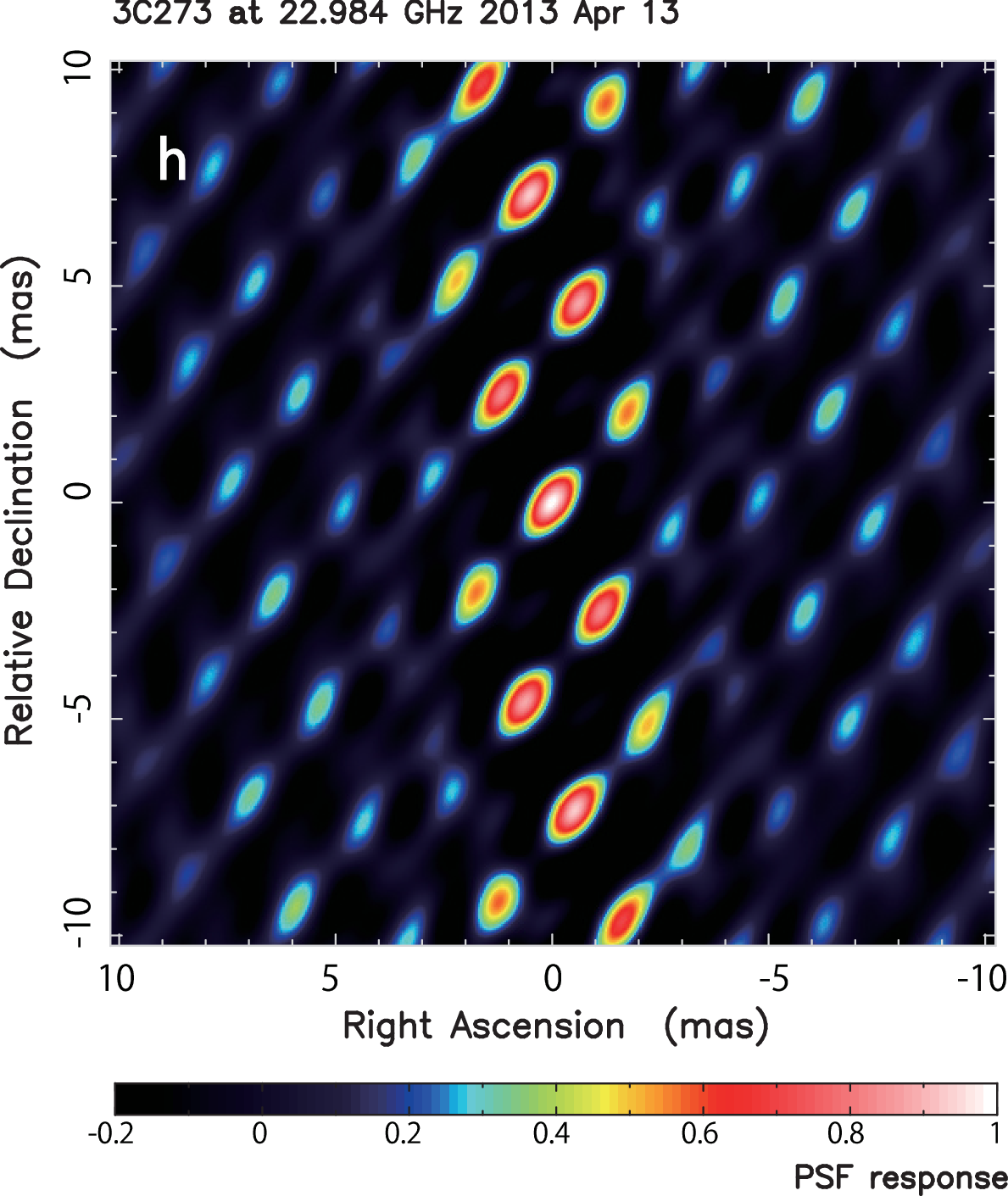}
\end{center}
\end{minipage}
\begin{minipage}{0.32\hsize}
\begin{center}
\includegraphics[width=\linewidth]{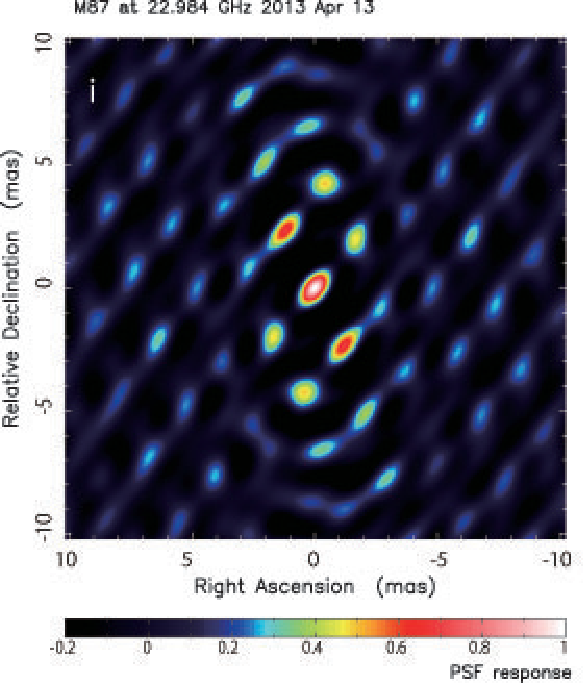}
\end{center}
\end{minipage}
\end{center}
\caption{\textit{uv}-coverage (\textit{top} panels) and point spread function of KaVA (\textit{middle} panels), VERA (\textit{bottom} panels) derived from the observation of 4C~39.25 (\textbf{a, d, g}), 3C~273 (\textbf{b, e, h}), and M~87 (\textbf{c, f, i}) at 23~GHz. In each \textit{uv}-coverage, red points derived solely from VERA observation. Observation was performed by 6-min $\times~11$ scans at an interval of ~30-min for during total observation time of 6-hours. The declination of each source is shown in the upper-right corner of each panel of \textit{uv}-coverage.} \label{fig:psf}
\end{figure*}

%
\subsection{4C~39.25 (J0927+3902)}
The high-frequency compact radio source 4C~39.25 located at a redshift of $z=0.695$ (Abazajian et al. 2004, 1~mas~$=7.1~\mathrm{pc}$) is known as a peculiar superluminal radio source \citep{marcaide85}. This source is reported to have a very bright jet structure compared with its radio core, and the brightness of the jet concentrates on the central 1~mas region on self-calibrated VLBI images at frequencies ranging from 15 to 43 GHz \citep{alberdi00}. 
Although the central 1 mas region of VLBI images acquired with insufficient angular resolution (e.g. $>~1~\mathrm{mas}$) has relatively compact structure, the submilliarcsecond structure of this source consists of a complex bent core-jet structure.

\subsection{3C~273 (J1229+0203)}
3C~273 is the first identified quasar \citep{schmidt63} classified as the flat-spectrum radio quasar (FSRQ) at a redshift of $z=0.158$ (Strauss et al. 1992, 1~mas~$=2.7~\mathrm{pc}$), and it is well known as one of the most powerful FSRQs.
Since the jet structure of this source has an extended and bright knotty structure in the south-west direction on the celestial sphere, a number of VLBI observations of 3C~273 have been conducted to study the nature of relativistic jet. Its superluminal motion was confirmed by VLBI images in the late 1970's \citep{pearson81}.

\subsection{M~87 (J1230+1223)}
M~87 is one of the most famous FR-I-type radio galaxies intensively studied by VLBI, because of its proximity ($z=0.00436$, Rines \& Geller 2008). At the distance of M~87, an angular resolution of 1 mas~=~0.08~pc corresponds to 140 Schwarzschild radii ($R_s$) for the mass of the central black hole, $M_{BH}~=~6.0\times10^9~M_{\solar}$~\citep{gebhardt09}. Although this source has a smoothed one-sided jet, the limb-brightened structure in its extended jet is well confirmed by many VLBI observations \citep[and references therein]{hada13}.

\begin{figure}[htbp]
\begin{center}

\includegraphics[width=0.95\linewidth]{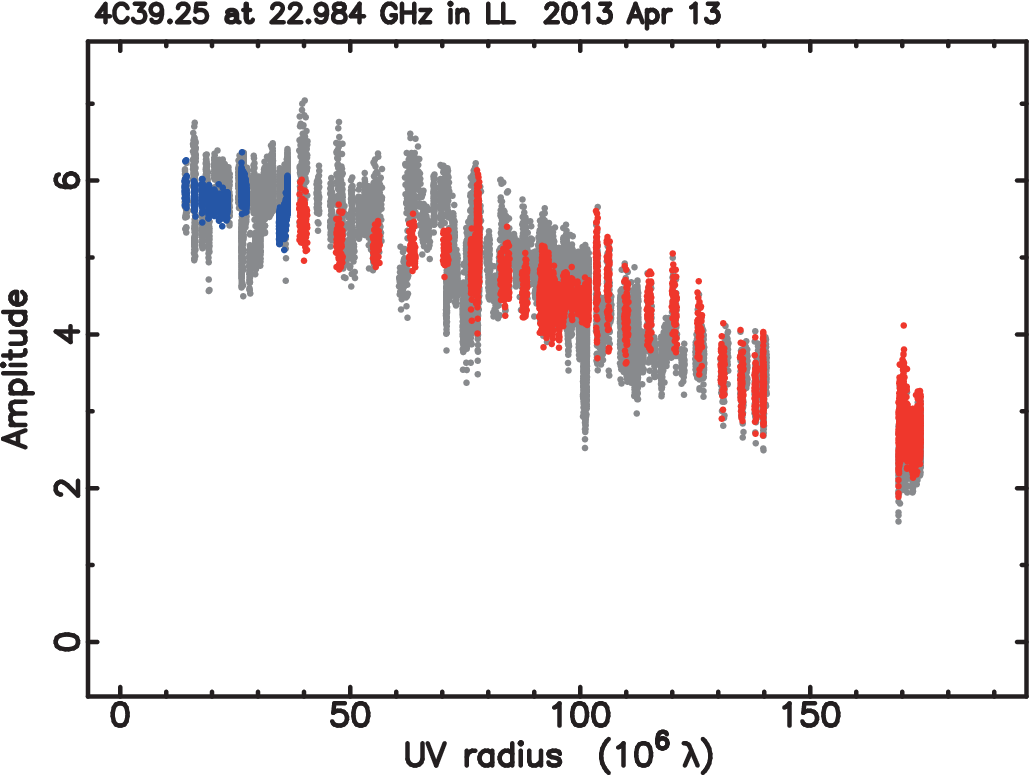}

\vspace{0.1in}

\includegraphics[width=0.95\linewidth]{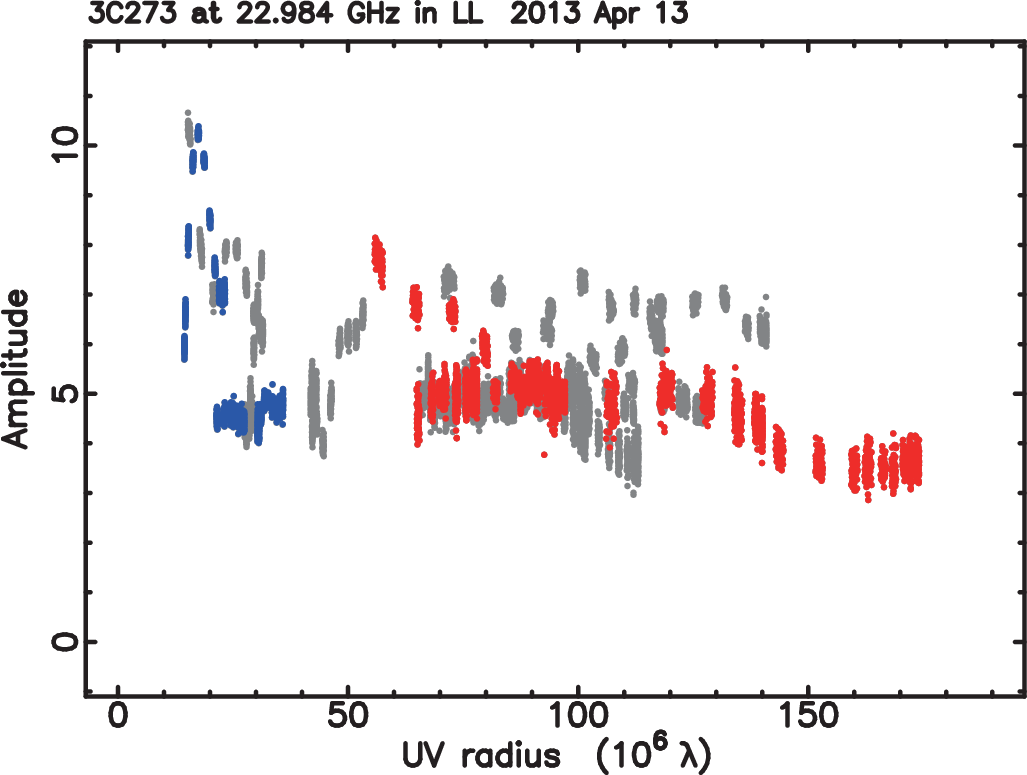}

\vspace{0.1in}

\includegraphics[width=0.95\linewidth]{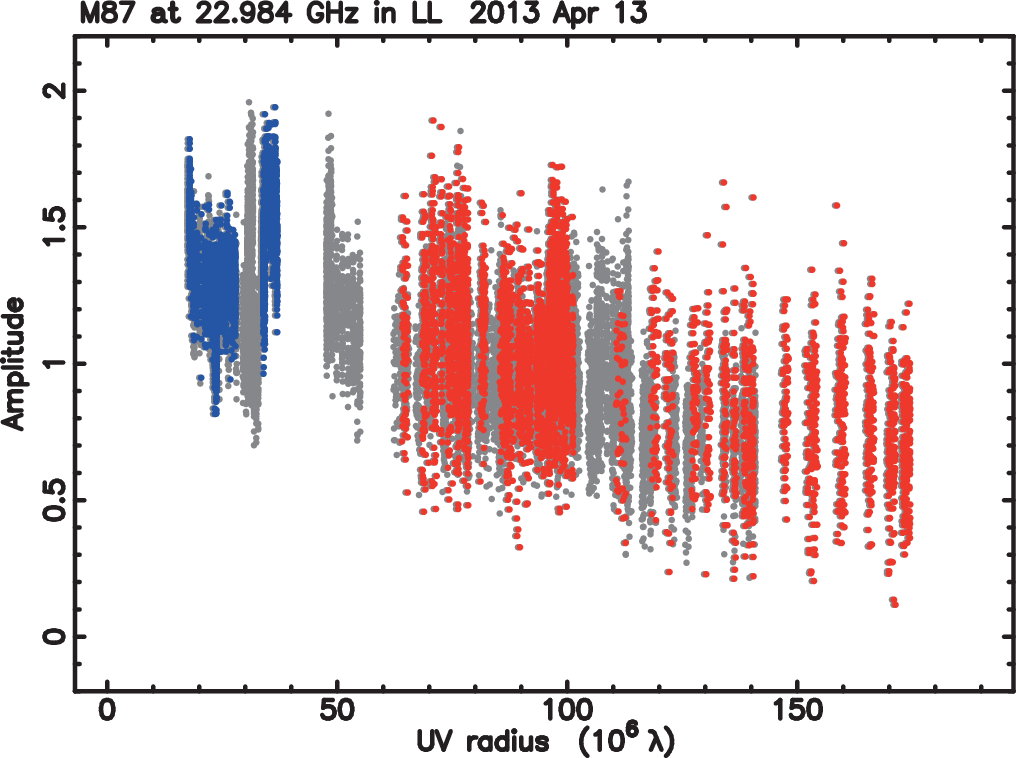}
\end{center}
\caption{Visibility amplitude plots of 4C~39.25 ({\it top}), 3C~273 ({\it middle}), and M~87 ({\it bottom}) along {\it uv}-distance at 23-GHz. Blue, red, and gray plots show the visibility derived from each individual array of KVN or VERA and from the combined array KaVA, respectively.}\label{fig:radplot}
\end{figure}

\section{KaVA observation and data reduction of bright AGN jets}\label{obs}
Our observations were made on April 13, 2013 at 23 GHz (code: r13103a) and on April 14, 2013 at 43 GHz (code: r13104a). All seven stations (three stations from KVN, and four stations from VERA) participated on both dates, and the total observation time was 6 h at 23 GHz, and 8 h at 43 GHz. To achieve good $uv$--coverages, we repeated 11 to 16 short-time scans (4 to 6 min/scan) for each source as follows: scan1 (source1) $\to$ scan2 (source2) $\to$ scan3 (source3) $\to$ scan4 (source1) $\to$ scan5 (source2) $\to$ scan6 (source3) $\to~\cdots$. Therefore, total on-source time for each source was approximately 1 h. 
As shown in Figure \ref{fig:psf}, KaVA observations clearly improved \textit{uv}-coverage and sidelobe levels compared with VERA observations, even though the declination of the source is relatively low ($\delta\sim2^{\circ}$).

Left-handed circular polarization was received and sampled with two-bit quantization, and filtered by using a digital filter system \citep{iguchi05, oh11}. The data were recorded onto a magnetic tape at a rate of 128 Mbps, providing a total bandwidth of 32 MHz that was divided into two intermediate frequency (IF) channels (16 MHz each). Correlation processing was performed on the Mitaka FX correlator \citep{chikada91} at the NAOJ Mitaka campus.

The calibration of correlated visibility was performed using the NRAO Astronomical Image Processing System (AIPS) software package \citep{greisen03}. First, we flagged the data of 10\% at both band edges in each IF, and then normalized the cross-correlation by the auto correlation. A standard {\it a priori} amplitude calibration was performed by the AIPS task APCAL, which is based on the measurements of the opacity-corrected system temperature ($T_\mathrm{sys}^*$) during observations and the antenna aperture efficiency for each station. 
The amplitude calibration error in both the VERA and KVN arrays has already been discussed by \citet{petrov12a, petrov12b}, respectively. In their works, the calibration error was estimated to be $\sim15\%$ for each array independently. 
In addition to {\it a priori} amplitude calibration, we also tried the "template" method at 43 GHz to confirm the best way to calibrate the amplitude of KaVA observations. The template method allows us to directly measure the system equivalent flux density (SEFD) in Jansky that is contemporaneous with data collection and corrects for antenna pointing errors by using the auto-correlation spectra on and off the line source (Reid 1995, Diamond 1995). For our observations, we derived the amplitude calibration table from the auto-correlation spectra of a SiO maser in R LMi using the AIPS task ACFIT. Next, we interpolated the calibration table to the continuum sources. 
Comparing the amplitude calibration tables derived from both the \textit{a priori} and the "template" methods, we found no significant difference compared with estimated calibration error between both methods ($<$5\%). Therefore, we conducted further analysis by using \textit{a priori} calibrated visibilities.
Fringe fitting was performed by the AIPS task FRING.
During 43 GHz observation session on April 14, 2013, the fringes based on the Iriki and the Ishigaki-jima stations were not detected for over half the observation time for all sources because of bad weather condition.

The calibrated visibility of each source was then exported to be imaged by the Caltech Difmap package \citep{shepherd97}. After flagging outliers of data, we time averaged the visibility data in 20 to 30 s/bin, depending on the signal-to-noise ratio (SNR) of the sources.
We show the visibility amplitude plots along {\it uv}-distance for each source in Figure \ref{fig:radplot}. In this figure, a significant difference in the calibrated visibility of 3C~273, which has a complex and extended jet structure, appears clearly between images acquired by individual arrays and by the combined array. 
As seen in Figures \ref{fig:psf} and \ref{fig:radplot}, in reproducing detailed jet structure, KaVA is more effective than the individual arrays at collecting the Fourier components with various spatial frequencies.
We iterated CLEAN and phase self-calibration algorithm by checking the consistency between the model and the observed closure phase and total flux until the residual error between the model and the observed visibility converged. And then, we repeated the phase and amplitude self-calibration until reasonable solutions were obtained. 
After the amplitude self-calibration, we evaluated whether systematic gain error appeared between KVN and VERA, except for the 43 GHz observation (because of its bad weather condition). We found that the averaged scaling factors of the antenna gain for each source are less than 10\% at 23 GHz, and the standard deviations among all seven stations are 4\%-7\% (see appendix for details). Therefore, we conclude that no significant systematic gain error occurs between KVN and VERA.

\section{Results}\label{result}

\subsection{KaVA images}

\begin{figure*}[htbp]

\begin{minipage}{0.32\hsize}
\begin{center}
\includegraphics[width=0.95\linewidth]{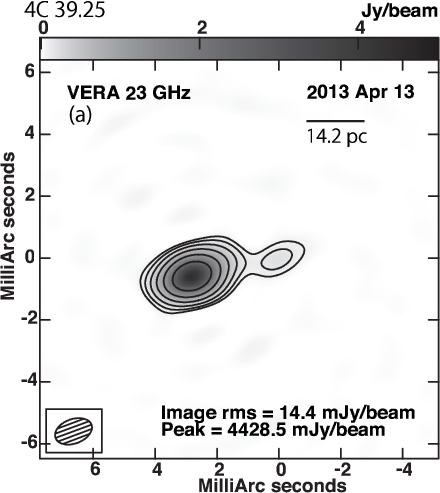}
\end{center}
\end{minipage}
\begin{minipage}{0.32\hsize}
\begin{center}
\includegraphics[width=0.95\linewidth]{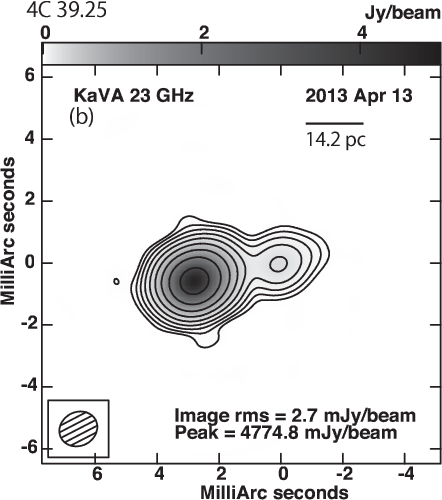}
\end{center}
\end{minipage}
\begin{minipage}{0.32\hsize}
\begin{center}
\includegraphics[width=0.95\linewidth]{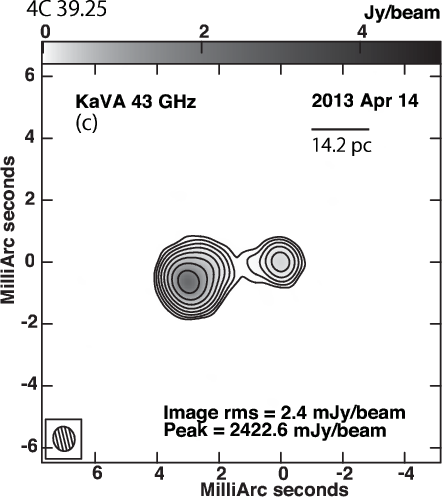}
\end{center}
\end{minipage}

\vspace{0.1in}

\begin{minipage}{0.32\hsize}
\begin{center}
\includegraphics[width=0.95\linewidth]{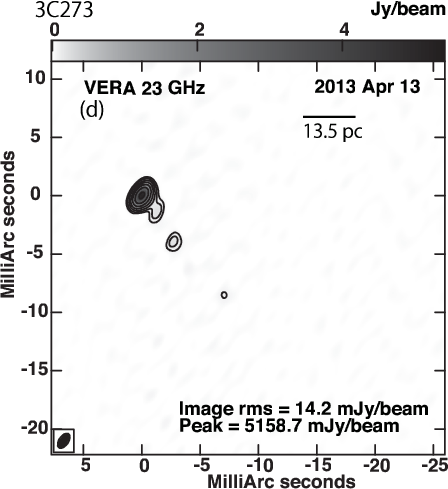}
\end{center}
\end{minipage}
\begin{minipage}{0.32\hsize}
\begin{center}
\includegraphics[width=0.95\linewidth]{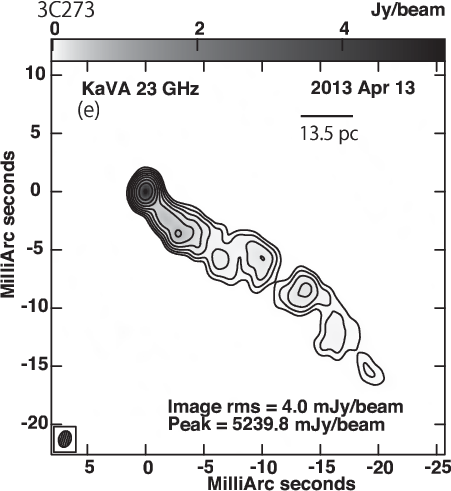}
\end{center}
\end{minipage}
\begin{minipage}{0.32\hsize}
\begin{center}
\includegraphics[width=0.95\linewidth]{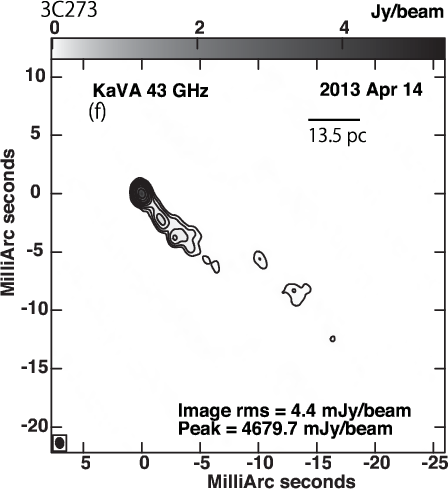}
\end{center}
\end{minipage}

\vspace{0.1in}

\begin{minipage}{0.32\hsize}
\begin{center}
\includegraphics[width=0.95\linewidth]{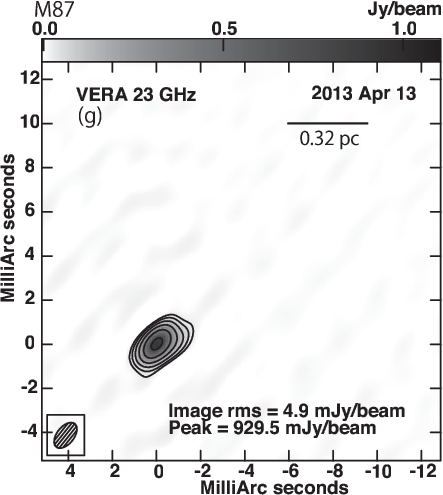}
\end{center}
\end{minipage}
\begin{minipage}{0.32\hsize}
\begin{center}
\includegraphics[width=0.95\linewidth]{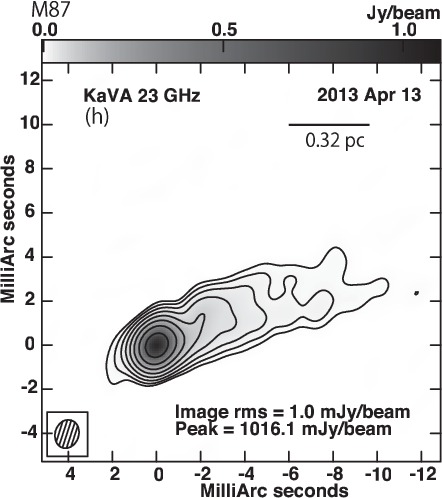}
\end{center}
\end{minipage}
\begin{minipage}{0.32\hsize}
\begin{center}
\includegraphics[width=0.95\linewidth]{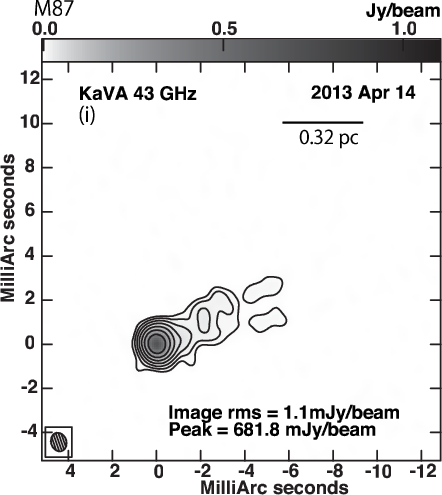}
\end{center}
\end{minipage}

\caption{(a) -- (c) VLBI images of 4C~39.25, (d) -- (f) 3C~273, and (g) -- (i) M~87 observed by VERA in the K band and KaVA in the K and Q bands, respectively. In each panel, peak intensity and image noise levels are indicated in the bottom-right corner and synthesized beams are also shown in the bottom-left corner in each image. In the top-right corner, the date and linear scale at the distance of each source are shown. Contours of each image begin at five times of image rms, and increase in $2^n$ steps.}\label{fig:image}
\end{figure*}

Figure. \ref{fig:image} presents the VLBI images of 4C~39.25, 3C~273, and M~87 observed by VERA at 23 GHz and by KaVA at 23 and 43 GHz with natural weighting.  
The images shown of VERA and KaVA are derived from the same data (observing code was r13103a at 23 GHz and r13104a at 43 GHz). We generated the VERA images by clipping the visibilities of VERA out of the AIPS-calibrated visibilities. Because the imaging capability of KVN was recently evaluated by \citet{lee14}, we focus on evaluating and comparing the imaging capabilities of KaVA and VERA in this paper.

To evaluate the quality of VLBI images, we follow \citet{lobanov06} and \citet{lee14} and estimate the quality of the residual noise distribution $\xi_\mathrm{r}$, which is expressed as $S_\mathrm{r}/S_\mathrm{r,exp}$ where $S_\mathrm{r}$ is the maximum absolute amplitude in the residual image and $S_\mathrm{r,exp}$ is the expected value of $S_\mathrm{r}$ from image noise level $\sigma_\mathrm{im}$ under the assumption of Gaussian noise with zero mean. 
When the residual noise approaches Gaussian noise, $\xi_\mathrm{r}\to1$. If $\xi_\mathrm{r}>1$, not all of the structure is adequately recovered; if $\xi_\mathrm{r}<1$, the image model has an excessively large number of degrees of freedom. 
The quality $\xi_\mathrm{r}$ of the images derived from KaVA range from 0.88 to 1.13. Since $S_\mathrm{r}$ and $S_\mathrm{r,exp}$ should also be affected by the amplitude calibration error mentioned in Section \ref{obs}, the parameter $\xi_\mathrm{r}$ has an uncertainty of approximately 20\%. Therefore, the result of this evaluation implies that KaVA images presented in this paper provide a good representation of the structures detected in the visibility data.

In Table \ref{tab:tbl2}, we present the peak intensity ($I_\mathrm{p}$), total flux ($S_\mathrm{total}$), $\sigma_\mathrm{im}$, $\xi_\mathrm{r}$, and synthesized beam size with natural weighting for each source. The structure of 4C~39.25, which has the core and bright compact jet structures, is well reproduced, even by only VERA, and there is no significant difference in peak intensity, total flux, or structure between the VERA and KaVA images. However, the image quality for 3C~273 and M~87, which have extended structures is clearly improved by using KaVA compared with using only VERA. By comparing KaVA images with VERA images, we find that the total fluxes in KaVA images are 30\%--40\% larger than those in VERA images. This difference implies that the extended emission of these two sources is well reproduced by KaVA.

\begin{figure}[htbp]
\centering\includegraphics[width=0.95\linewidth]{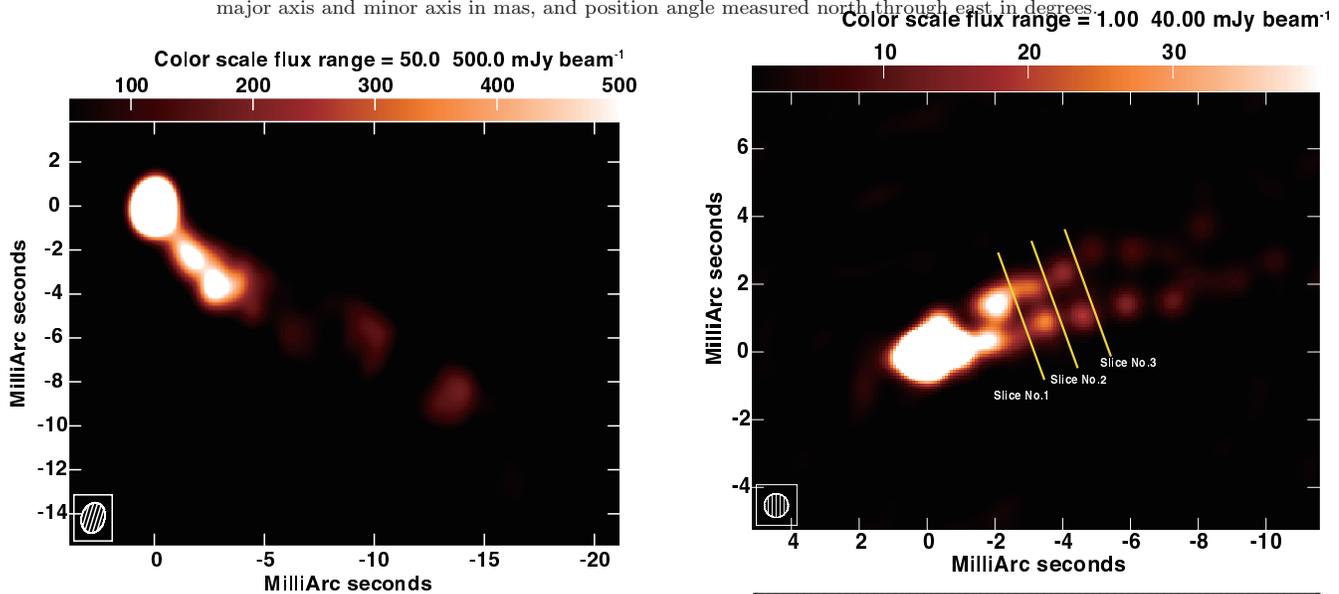}
\caption{Same image as Figure \ref{fig:image}-(e) but substructures in the extended emission are emphasized by different color contrast.}\label{fig:knotty}
\end{figure}

\begin{figure}[htbp]
\centering
\includegraphics[width=0.95\linewidth]{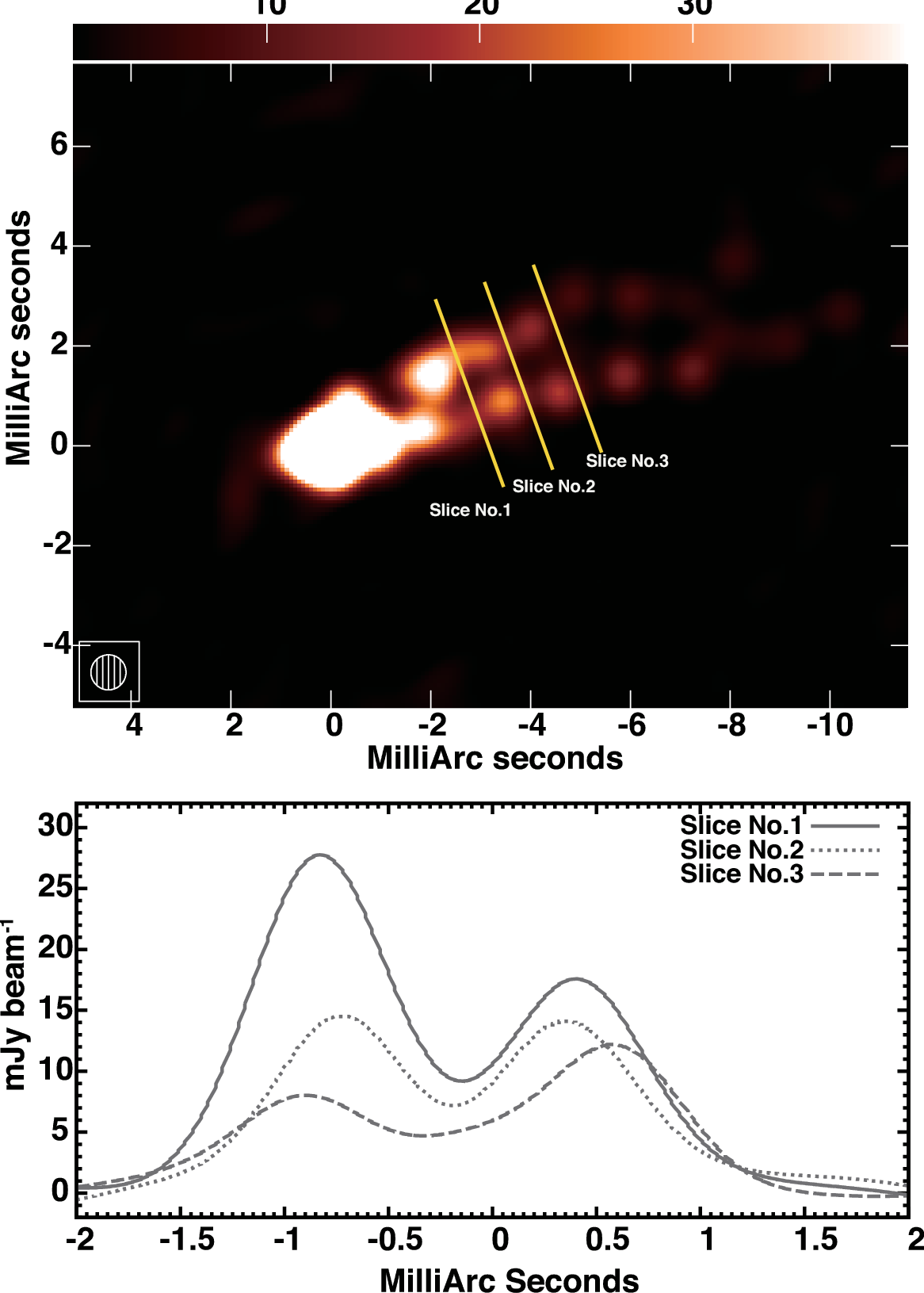}
\caption{KaVA image of M~87 at 23-GHz (\textit{upper} panel) and the transverse brightness across M87 jet (\textit{lower} panel). The \textit{upper} panel is the same image as in Figure \ref{fig:image}-(h) but is convolved with a circular beam size of 0.7~mas (typical synthesized beam size at 43 GHz). The horizontal axis represents relative distance along each yellow line shown in the \textit{upper} panel. The yellow lines show the direction perpendicular to the jet axis (PA of $-70^{\circ}$) at 3, 4, and 5 mas from the image center along the jet.}\label{fig:limb}
\end{figure}

In addition, as shown in Figures \ref{fig:image}-(e), \ref{fig:image}-(f), \ref{fig:image}-(h) and \ref{fig:image}-(i), we find that KaVA clearly recovers not only the extended jet but also the well-known characteristic structures, bright knotty substructures of the jet in 3C~273 and limb-brightened of the smoothed jet in M~87 at both 23 and 43 GHz. 
In Figures \ref{fig:knotty} and \ref{fig:limb}, we show the same 23 GHz image of 3C~273 and M~87 as shown in Figures \ref{fig:image}-(e), \ref{fig:image}-(h), respectively, but those characteristic structures are pronounced. For 3C~273, although VERA could not recover the extended emission, the bright knotty substructures were well reproduced, as shown in Figure \ref{fig:image}-(d).

\begin{table*}[htbp]
  \centering
  \caption{Image parameters.}\label{tab:tbl2}
    \scalebox{0.95}{\begin{tabular}{llccccc}
    \toprule\toprule
    Array (Band) & Source & $I_\mathrm{p}$ & $S_\mathrm{total}$ & $\sigma_\mathrm{im}$ & $\xi_\mathrm{r} $ & $\theta_\mathrm{maj}\times\theta_\mathrm{min}$ (P.A.)\\\addlinespace[0.05in]
          &       &   (1)    &    (2)   &    (3)   &    (4)   &  (5) \\\midrule
    VERA (K) & 4C~39.25 & 4428.5 & 6062.6 & 14.4  &   1.29    & $1.24\times0.79~(-69.6)$\\\addlinespace[0.05in]
          & 3C~273 & 5158.7  & 6486.9 & 14.2   &    1.24   & $1.42\times0.78~(-34.3)$\\\addlinespace[0.05in]
          & M~87  & 929.5 & 1037.4  & 4.9   &    1.66   & $1.34\times0.78~(-38.6)$ \\\addlinespace[0.1in]
    KaVA (K) & 4C~39.25 & 4774.8 & 6371.2 & 2.7   &   0.88    & $1.26\times1.09~(-62.8)$ \\\addlinespace[0.05in]
          & 3C~273 & 5239.8 & 10457.1 & 4.0   &    1.13   & $1.40\times1.04~(-17.0)$ \\\addlinespace[0.05in]
          & M~87  & 1016.9 & 1669.4 & 1.0   &    1.10   & $1.30\times1.07~(-17.9)$ \\\addlinespace[0.1in]
    KaVA (Q) & 4C~39.25 & 2422.6 & 4087.2 & 2.4   &    1.05   & $0.78\times0.68~(19.7)$ \\\addlinespace[0.05in]
          & 3C~273 & 4679.7 & 6818.0 & 4.4   &   1.09    & $0.86\times0.65~(7.9)$ \\\addlinespace[0.05in]
          & M~87  & 681.8 & 1082.4 & 1.1   &   1.13    & $0.85\times0.69~(19.2)$ \\\bottomrule
\multicolumn{7}{@{}l@{}}{\hbox to 0pt{\parbox{130mm}{\footnotesize
\textbf{Notes.~}(1): Peak intensity in \mjybm. (2): Total flux density in mJy. (3): Image rms noise level in \mjybm. (4): Quality of the residual noise in the image. (5): Synthesized beam size: major axis and minor axis in mas, and position angle measured north through east in degrees.
}\hss}}
    \end{tabular}}%
\end{table*}

\subsection{Model-fit images}

\begin{figure*}[htbp]
\centering
\includegraphics[width=0.95\linewidth]{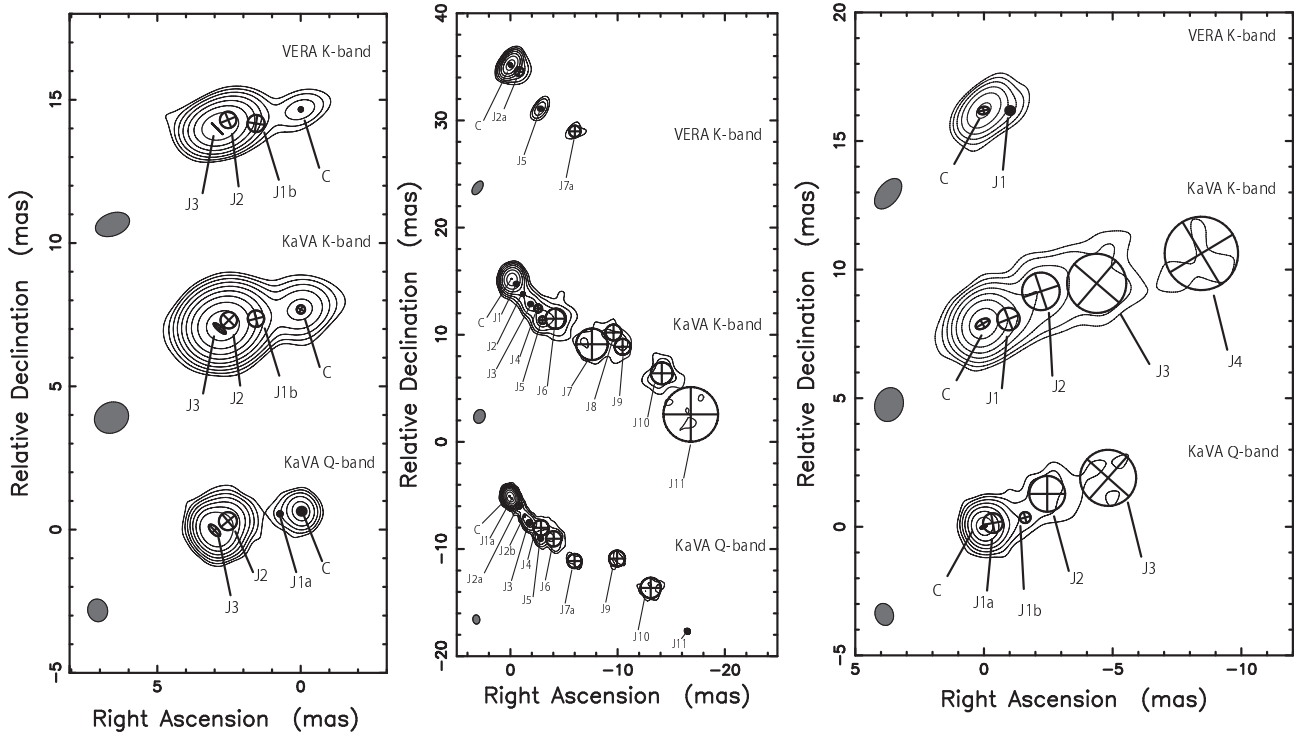}
\caption{Model fit images of 4C~39.25 (\textit{left}), 3C~273 (\textit{center}), and M~87 (\textit{right}) with natural weighting. The labels "C" and "J" represent the radio core, and jet component, respectively. For the jet components, we labeled the number together with "J" in order of distance from the radio core "C". Image contour begin at five times the image rms and increase in $2^n$ steps.}\label{fig:mdfit}
\end{figure*}

To confirm how well the extended jet structures are reproduced by using KaVA compared with by using VERA, we fit two-dimensional Gaussian models to visibility data. 
 In this procedure, we applied the elliptical or the circular Gaussian model to the self-calibrated visibility data in the $uv$-plane by using the Difmap task {\it modelfit}.
Here, we selected a suitable Gaussian model for each jet component by checking the consistency of the characteristic structures between model-fit images and total intensity images created by CLEAN at 23 and 43 GHz, and the goodness of the fit was judged by relative $\chi^2$ statistics.
In Figure \ref{fig:mdfit}, we show the model-fit images of VERA at 23~GHz, and of KaVA at 23 and 43-GHz for each source. In addition, we list the parameters of each Gaussian model in Table \ref{tab:tbl3}. The uncertainties of each parameter are estimated by following \citet[and references therein]{lee08}. These uncertainties are possibly underestimated because systematic errors (e.g., derived from instruments) are not considered. Their evaluation is left as a future task.

\begin{table*}[htbp]
  \centering
  \caption{parameters for Gaussian model.}\label{tab:tbl3}
    \scalebox{0.7}{\begin{tabular}{lcccccccc}
    \toprule\toprule
    Source & Array (Band) & ID    & $I_\mathrm{p}$ & $S$ & $r$   & $\theta$ & $a~(b)$ & $\phi$ \\\addlinespace[0.05in]
          & (1) & (2) & (3) & (4) & (5) & (6) & (7) & (8) \\
    \midrule
    4C~39.25 & VERA (K) & C     & $222\pm35$ & $221\pm49$ & $\cdots$ & $\cdots$ & $<0.17$ &  \\\addlinespace[0.05in]
          &       & J1b   & $80\pm17$ & $108\pm28$ & $1.64\pm0.06$ & $107.4\pm2.2$ & $0.60\pm0.13$ &  \\\addlinespace[0.05in]
          &       & J2    & $1491\pm94$ & $2017\pm158$ & $2.56\pm0.02$ & $98.1\pm0.4$ & $0.58\pm0.04$ &  \\\addlinespace[0.05in]
          &       & J3    & $3243\pm120$ & $3836\pm185$ & $2.99\pm0.01$ & $103.0\pm0.2$ & $0.54\pm0.02 (<0.04)$ & $44.9\pm2.1$ \\\addlinespace[0.1in]
          & KaVA (K) & C     & $238\pm18$ & $251\pm26$ & $\cdots$ & $\cdots$ & $0.30\pm0.02$ &  \\\addlinespace[0.05in]
          &       & J1b   & $133\pm10$ & $164\pm16$ & $1.64\pm0.02$ & $101.6\pm0.8$ & $0.59\pm0.05$ &  \\\addlinespace[0.05in]
          &       & J2    & $1904\pm53$ & $2342\pm85$ & $2.61\pm0.01$ & $98.5\pm0.2$ & $0.59\pm0.02$ &  \\\addlinespace[0.05in]
          &       & J3    & $3208\pm69$ & $3583\pm104$ & $2.98\pm0.01$ & $103.2\pm0.1$ & $0.56\pm0.01 (0.17\pm0.01)$ & $49.6\pm1.0$ \\\addlinespace[0.1in]
          & KaVA (Q) & C     & $298\pm15$ & $294\pm22$ & $\cdots$ & $\cdots$ & $<0.06$ &  \\\addlinespace[0.05in]
          &       & J1a   & $29\pm6$ & $47\pm11$ & $0.75\pm0.02$ & $96.8\pm1.4$ & $0.18\pm0.04$ &  \\\addlinespace[0.05in]
          &       & J2    & $600\pm50$ & $718\pm79$ & $2.58\pm0.03$ & $97.7\pm0.6$ & $0.63\pm0.05$ &  \\\addlinespace[0.05in]
          &       & J3    & $2188\pm96$ & $3009\pm163$ & $3.10\pm0.01$ & $102.4\pm0.2$ & $0.52\pm0.02 (0.24\pm0.02)$ & $44.0\pm2.2$ \\\midrule
    3C~273 & VERA (K) & C     & $5152\pm99$ & $5951\pm151$ & $\cdots$ & $\cdots$ & $0.45\pm0.01 (0.27\pm0.01)$ & $36.4\pm1.3$ \\\addlinespace[0.05in]
          &       & J2a    & $317\pm36$ & $557\pm73$ & $1.00\pm0.05$ & $-127.2\pm2.7$ & $0.82\pm0.09$ &  \\\addlinespace[0.05in]
          &       & J5    & $421\pm55$ & $414\pm77$ & $4.80\pm0.08$ & $-145.9\pm0.5$ & $<0.17$ &  \\\addlinespace[0.05in]
          &       & J7a    & $85\pm28$ & $199\pm72$ & $8.41\pm0.17$ & $-135.9\pm1.2$ & $1.05\pm0.35$ &  \\\addlinespace[0.1in]
          & KaVA (K) & C     & $4727\pm119$ & $4765\pm169$ & $\cdots$ & $\cdots$ & $0.11\pm0.00$ &  \\\addlinespace[0.05in]
          &       & J1    & $1034\pm34$ & $1036\pm48$ & $0.60\pm0.06$ & $-138.9\pm2.6$ & $<0.11$ &  \\\addlinespace[0.05in]
          &       & J2    & $334\pm22$ & $354\pm33$ & $1.72\pm0.01$ & $-142.5\pm0.4$ & $0.32\pm0.02$ &  \\\addlinespace[0.05in]
          &       & J3    & $521\pm42$ & $525\pm60$ & $2.85\pm0.08$ & $-143.5\pm0.8$ & $<0.16$ &  \\\addlinespace[0.05in]
          &       & J4    & $209\pm31$ & $272\pm51$ & $3.63\pm0.05$ & $-137.9\pm0.8$ & $0.71\pm0.10$ &  \\\addlinespace[0.05in]
          &       & J5    & $531\pm36$ & $692\pm59$ & $4.70\pm0.02$ & $-143.1\pm0.3$ & $0.69\pm0.05$ &  \\\addlinespace[0.05in]
          &       & J6    & $318\pm58$ & $1022\pm197$ & $5.41\pm0.17$ & $-132.0\pm1.8$ & $1.85\pm0.34$ &  \\\addlinespace[0.05in]
          &       & J7    & $127\pm42$ & $635\pm214$ & $9.43\pm0.47$ & $-129.1\pm2.9$ & $2.87\pm0.95$ &  \\\addlinespace[0.05in]
          &       & J8    & $72\pm26$ & $185\pm72$ & $10.48\pm0.26$ & $-117.7\pm1.4$ & $1.44\pm0.52$ &  \\\addlinespace[0.05in]
          &       & J9    & $112\pm31$ & $261\pm79$ & $11.85\pm0.20$ & $-121.4\pm1.0$ & $1.46\pm0.40$ &  \\\addlinespace[0.05in]
          &       & J10   & $196\pm57$ & $697\pm210$ & $16.15\pm0.29$ & $-122.1\pm1.0$ & $1.99\pm0.58$ &  \\\addlinespace[0.05in]
          &       & J11   & $75\pm29$ & $906\pm351$ & $20.45\pm0.96$ & $-127.1\pm2.7$ & $4.98\pm1.92$ &  \\\addlinespace[0.1in]
          & KaVA (Q) & C     & $3274\pm68$ & $3368\pm97$ & $\cdots$ & $\cdots$ & $0.13\pm0.00$ &  \\\addlinespace[0.05in]
          &       & J1a   & $1836\pm58$ & $1883\pm82$ & $0.31\pm0.00$ & $-139.8\pm0.3$ & $0.12\pm0.00$ &  \\\addlinespace[0.05in]
          &       & J2a   & $131\pm24$ & $202\pm44$ & $1.23\pm0.05$ & $-136.9\pm2.5$ & $0.57\pm0.11$ &  \\\addlinespace[0.05in]
          &       & J2b   & $142\pm16$ & $146\pm23$ & $2.35\pm0.01$ & $-143.6\pm0.2$ & $0.17\pm0.02$ &  \\\addlinespace[0.05in]
          &       & J3    & $205\pm24$ & $274\pm40$ & $3.07\pm0.03$ & $-143.7\pm0.5$ & $0.46\pm0.05$ &  \\\addlinespace[0.05in]
          &       & J4    & $82\pm25$ & $277\pm87$ & $4.16\pm0.22$ & $-135.4\pm3.0$ & $1.45\pm0.44$ &  \\\addlinespace[0.05in]
          &       & J5    & $160\pm27$ & $155\pm38$ & $4.88\pm0.05$ & $-144.5\pm0.3$ & $<0.11$ &  \\\addlinespace[0.05in]
          &       & J6    & $113\pm32$ & $359\pm107$ & $5.72\pm0.20$ & $-134.1\pm2.0$ & $1.43\pm0.41$ &  \\\addlinespace[0.05in]
          &       & J7a   & $39\pm15$ & $108\pm45$ & $8.56\pm0.26$ & $-135.0\pm1.8$ & $1.33\pm0.53$ &  \\\addlinespace[0.05in]
          &       & J9    & $46\pm18$ & $127\pm52$ & $11.58\pm0.29$ & $-120.2\pm1.4$ & $1.50\pm0.58$ &  \\\addlinespace[0.05in]
          &       & J10   & $57\pm18$ & $234\pm76$ & $15.69\pm0.30$ & $-123.1\pm1.1$ & $1.91\pm0.60$ &  \\\addlinespace[0.05in]
          &       & J11   & $26\pm6$ & $31\pm10$ & $20.82\pm0.06$ & $-127.3\pm0.2$ & $0.45\pm0.11$ &  \\\midrule
    M~87  & VERA (K) & C     & $933\pm43$ & $1049\pm65$ & $\cdots$ & $\cdots$ & $0.42\pm0.02 (0.28\pm0.02)$ & $-78.6\pm2.7$ \\\addlinespace[0.05in]
          &       & J1    & $9\pm5$ & $93\pm49$ & $0.99\pm0.43$ & $-89.9\pm12.2$ & $<0.85$ &  \\\addlinespace[0.1in]
          & KaVA (K) & C     & $992\pm32$ & $1131\pm48$ & $\cdots$ & $\cdots$ & $0.58\pm0.02 (0.33\pm0.02)$ & $-60.1\pm2.0$ \\\addlinespace[0.05in]
          &       & J1    & $119\pm13$ & $180\pm24$ & $0.99\pm0.05$ & $-79.5\pm2.8$ & $0.85\pm0.10$ &  \\\addlinespace[0.05in]
          &       & J2    & $67\pm16$ & $157\pm40$ & $2.53\pm0.17$ & $-60.8\pm3.8$ & $1.42\pm0.33$ &  \\\addlinespace[0.05in]
          &       & J3    & $32\pm11$ & $141\pm49$ & $4.58\pm0.38$ & $-70.4\pm4.7$ & $2.24\pm0.75$ &  \\\addlinespace[0.05in]
          &       & J4    & $16\pm8$ & $77\pm37$ & $8.70\pm0.65$ & $-72.1\pm4.3$ & $2.78\pm1.30$ &  \\\addlinespace[0.1in]
          & KaVA (Q) & C     & $590\pm20$ & $630\pm30$ & $\cdots$ & $\cdots$ & $0.26\pm0.01 (0.10\pm0.01)$ & $-63.5\pm2.2$ \\\addlinespace[0.05in]
          &       & J1a   & $149\pm10$ & $266\pm21$ & $0.38\pm0.03$ & $-68.5\pm3.9$ & $0.75\pm0.05$ &  \\\addlinespace[0.05in]
          &       & J1b   & $20\pm4$ & $25\pm6$ & $1.61\pm0.04$ & $-77.5\pm1.5$ & $0.42\pm0.08$ &  \\\addlinespace[0.05in]
          &       & J2    & $34\pm9$ & $98\pm28$ & $2.71\pm0.19$ & $-62.6\pm3.9$ & $1.36\pm0.37$ &  \\\addlinespace[0.05in]
          &       & J3    & $13\pm5$ & $62\pm26$ & $5.07\pm0.44$ & $-68.6\pm5.0$ & $2.13\pm0.88$ &  \\
    \bottomrule\addlinespace[0.05in]
\multicolumn{8}{@{}l@{}}{\hbox to 0pt{\parbox{200mm}{\footnotesize
\textbf{Notes.~}(1): Array name, and observation frequency. K: 23 GHz, Q: 43 GHz. (2): The labels "C" and "J" represent the radio core, and the jet component, respectively. (3): Peak intensity of each component in \mjybm. (4): Flux density of each component in mJy. (5): Radial angular distance of each jet component relative to the core in mas. (6): Position angle of each jet component measured from north to east relative to the core in degrees. (7): Size of Gaussian models in mas. The value of ($b$) represents the minor axis size of models if elliptical Gaussian is used for model-fit. (8): Position angle of major-axis of elliptical Gaussian model measured from north to east in degrees.
}\hss}}
     \end{tabular}}
\end{table*}

\subsection{Spectral index maps}

\begin{figure*}[htbp]
\centering
\includegraphics[width=0.95\linewidth]{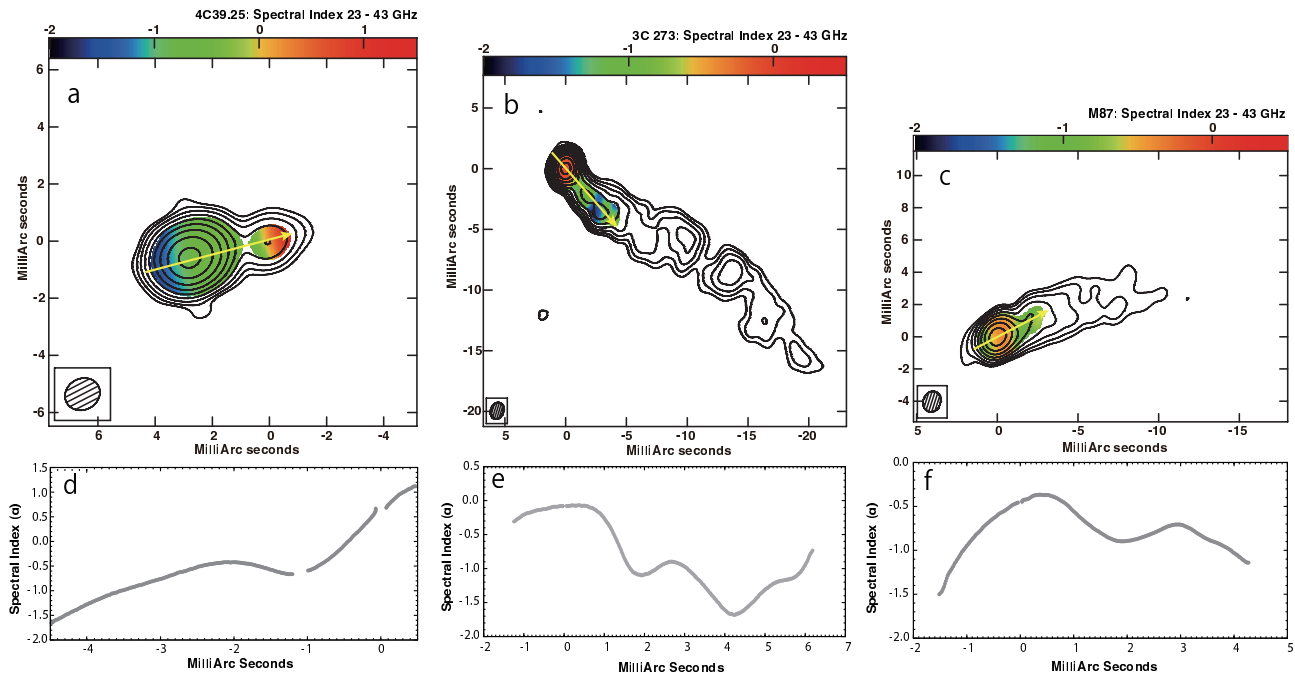}
\caption{(a) -- (c) Spectral index maps for 4C~39.25, 3C~273, and M~87, respectively, derived from flux densities at 23 and 43 GHz after position alignment. The flux density distribution at 23 GHz is shown as solid white contours without position alignment. Spectral index scale ranges from -2.0 to 1.5 for 4C~39.25 and from -2.0 to 0.5 for 3C~273, and M~87. Contours starts at 15.0\mjybm for 4C~39.25 and 3C~273 and at 5.0\mjybm for M~87, and increases by factors of 2.
To obtain the spectral index map, we set a flux cutoff of 100 mJy for calculating the spectral index of 4C~39.25 and 3C~273 and 30 mJy for that of M~87. The 43 GHz images were restored with a beam of 23 GHz [lower-left corners of panels (a) -- (c)]. 
(d) -- (f) Spectral index slices along the direction indicated by yellow arrows in each spectral index map. Horizontal axes show the distance from image center toward each arrow.
}\label{fig:spimap}
\end{figure*}

When we discuss the spectral indices of AGN jets, we have to consider (1) the variability of intensity and structures between the observations at both frequencies, (2) the core shift effect of the radio core, which must be optically thick (optical depth $\tau\sim1$) against synchrotron self-absorption (SSA; Lobanov 1998) causes positional misalignment of the radio core on the VLBI images between different frequencies. 
%
%

To calculate accurate spectral indices of target sources, their variation in intensity between observations at different frequencies must be negligible. To resolve this problem, it is essential to observe the target sources simultaneously or quasi-simultaneously at dual (or multiple) frequencies. 
In addition, the similar \textit{uv}-coverage and the same angular resolution between different frequencies allow us to compare source structures with the same sensitivity for extended structures. Therefore we conducted KaVA observations for these sources at 23 and 43 GHz within two days of each other and during almost the same hour angle, and restored the synthesized beam in the imaging process to match angular resolutions in both frequencies.
%
%

The positional misalignment of multiple-frequency VLBI images may also cause misidentification or incomplete distribution of spectral indices. To obtain an accurate distribution of spectral indices, we also have to align the position of VLBI images at different frequencies before calculating the spectral index distributions. However, self-calibrated VLBI images basically lack absolute position. Therefore, to align images acquired at different frequencies, we assume that the positions of optically thin components at different frequencies should be fixed on the celestial sphere. 
To align the position of the radio core between the images at 23 and 43 GHz, we calculated the average position shift by assuming fixed celestial components between 23 and 43 GHz (Table \ref{tab:tbl3}) for the J2 and J3 components of 4C~39.25, the J3, J5, and J6 components of 3C~273, and the J2 and J3 components of M~87. Finally, in 23 GHz images, we shifted the position ($\Delta\alpha,~\Delta\delta$) in mas of (0.05, 0.03) for 4C~39.25, (-0.07, -0.25) for 3C~273, and (-0.30, 0.17) for M~87 are performed respectively. Since the amount of these shifts in position are on the order of tenths to one half of the size of the synthesized beam for our observation at 43 GHz, they cannot be neglected. 
However, as shown in Table \ref{tab:tbl3}, the uncertainties in position for the jet components used for positional alignment are significantly large, as are the shifts in position especially for 3C~273 and M~87. 
For example, \citet{hada11} made quasi-simultaneous multi-$\nu$ astrometric VLBI observations of M~87, and they found a core shift of $< 0.1$ mas between 24 and 43 GHz. The core shift measured by using optically thin components is evidently large compared with that reported by \citet{hada11}. Therefore, we must account for uncertainties in the alignment of position in the image that are as large as the error in position of each jet component. 
After aligning the position, we use the AIPS task COMB to generate accurate spectral index ($\alpha$) maps between 23 and 43 GHz for each source (Figure \ref{fig:spimap}), with $\alpha=(\log S_{\nu2}-\log S_{\nu1})/(\log\nu2-\log\nu1)$~where $S_\nu$ is the observed flux density at frequency $\nu$ (i.e., $S_\nu\propto\nu^\alpha$).

The structure of 4C~39.25 consists roughly of a bright component with a steep spectrum $\alpha~<~-0.5$ located in the central 1 mas region on the self-calibrated VLBI image and 
a component with an optically thick inverted spectrum of $\alpha\sim1$ between 23 and 43 GHz, located at 2.8 mas west from the brightest component (Figure \ref{fig:spimap}-a). This optically thick component is thought to be the radio core of 4C~39.25 \citep{alberdi97}. \citet{alberdi97} suggested a spectral turnover frequency for this component in the range of 43 to 86 GHz. In addition, the stationarity of this core was also verified by phase-referencing observations \citep{guirado95}.
Therefore, based on the result presented in Section 5.2, we shifted the position of the radio core to the center of the images in Figures \ref{fig:image}-a, \ref{fig:image}-b, \ref{fig:image}-c, Figure \ref{fig:mdfit}-\textit{left}, and Figure \ref{fig:spimap}.
The flat-spectrum radio quasar 3C~273 also has an optically thick core component whose spectral index is relatively flat ($\alpha\sim0$) between 23 and 43 GHz [Figures \ref{fig:spimap}-b, \ref{fig:spimap}-e] and that is located at the center of the VLBI image.
This result is consistent with that previously reported by \citet{mantovani00}.

In contrast, despite detecting a core shift, the radio core in M87 observed by KaVA has a the steep spectrum [$\alpha\sim-0.4$; Figure \ref{fig:spimap}-c]. However, Figure \ref{fig:spimap}-f shows a spectral gradient along the jet, and opacity increases toward the core. In addition, it is known that the spectrum of M87 core derived from VLBA observation shows relatively flat or inverted between 23 and 43 GHz~\citep{hada12} as well as 4C~39.25 and 3C~273 observed by KaVA.
The M 87 core observed by KaVA has a steep spectrum probably because of the contribution of the optically thin jet over the SSA core, which is within the resolution of KaVA. In fact, this is consistent with what was detected by VLBA in the central 1.2 mas region \citep{hada12}, which is comparable to the size of KaVAfs synthesized beam for observation in the 23 GHz band.


\section{Evaluation of KaVA imaging capability}\label{evaluation}

In the radio interferometer, the detection limit (thermal noise level) $\sigma_\mathrm{th}$ of image
 is derived from following equation \citep{thompson01};
\begin{eqnarray}
\sigma_\mathrm{th}&=&\frac{2k_\mathrm{B}T_S}
{A\eta_Q\sqrt{n_a(n_a-1)\Delta\nu_\mathrm{IF}\tau_0}}
\end{eqnarray}
where, $k_\mathrm{B}, T_S, A, n_a, \eta_Q, \Delta\nu_\mathrm{IF}, \mathrm{and}~\tau_0$, are Boltzmann's constant, the geometric mean of a system temperature, and an efficient aperture area of participated antennas, the number of antennas, an efficiency factor related to quantization, bandwidth, and total integration time, respectively. Here, we consider the use of {\it natural weighting}. 
In our observations, $\sigma_{\mathrm{th, KaVA}}$ is expected to be 0.7 \mjybm at 23 GHz, and 1.7 \mjybm at 43 GHz. For each antenna, the quantities $A$ and $T_S$ used for the calculation were derived from the KaVA status report and the average system temperature during the observation, respectively. 
During our 43 GHz observation, the weather condition in VERA Iriki and VERA Ishigaki-jima stations were significantly worse ($T_S$ averages are $\sim2000~\mathrm{K}$) compared with the other five stations or 23 GHz observations. Without the Iriki and Ishigaki-jima stations, the calculated noise level is 1.1 \mjybm,~which is much better than that estimated by using all seven stations. In analyzing the data, we flagged out more than half of the scans in the data at 43 GHz from these two stations.

However, in practice, the quality of most VLBI images is limited by its dynamic rage. Nevertheless, it is difficult to formulate the dynamic range of each observation. 
Incomplete \textit{uv}-coverage and calibration errors of phase and/or amplitude in visibility data cause higher sidelobes in the point-spread function. Therefore, we consider this to be the cause of the systematic error, which leads a decrease in the dynamic range of VLBI images. Because brighter sources are much more affected by this systematic error, the noise level in these VLBI images is expected to be larger than the thermal noise.

To evaluate the dynamic range of radio interferometry images, \citet{perley99} derived the following relationship between the dynamic range ($D$) of images and the phase calibration error ($\Delta\phi$) or amplitude calibration error ($\epsilon$):
\begin{eqnarray}
D = \frac{\sqrt{M}\sqrt{N(N-1)}}{\Delta\phi},~D = \frac{\sqrt{M}\sqrt{N(N-1)}}{\epsilon}
\end{eqnarray}
where, $M$ is the number of scans and $N$ is the number of antennas in the observation.

Following the way discussed in \citet{perley99}, we compare the dynamic ranges in KaVA observations with those expected. If we assume no significant differences in phase- and amplitude-calibration errors between KVN and VERA baselines, the dynamic range of KaVA images is expected to be 2 to 3 times greater than that of VERA or KVN images. 
\citet{lee14} reported a dynamic range of 317 to 420 for KVN images acquired by multiple-snapshot observation ($M: 10-13$), which is similar to the way we made the observations presented in this paper. Therefore, for observations of brighter sources, we expect KaVA to achieve a dynamic range of at least $\sim1000$.

\begin{table}[htbp]
  \caption{Comparison of image dynamic range and SNR between arrays.}
  \centering
    \scalebox{0.75}{\begin{tabular}{lcccccccc}\toprule\toprule
    Source & \multicolumn{2}{c}{KaVA K-band} &       & \multicolumn{2}{c}{KaVA Q-band} &       & \multicolumn{2}{c}{VERA K-band}\\\cmidrule{2-3}\cmidrule{5-6}\cmidrule{8-9}
          & $I_\mathrm{p}/\sigma_\mathrm{im}$  & $I_\mathrm{p}/\sigma_\mathrm{th}$  &       & $I_\mathrm{p}/\sigma_\mathrm{im}$  & $I_\mathrm{p}/\sigma_\mathrm{th}$  &         & $I_\mathrm{p}/\sigma_\mathrm{im}$  & $I_\mathrm{p}/\sigma_\mathrm{th}$\\\midrule
    4C~39.25 & 1768  & 6821  &       & 1009  & 2202  &       & 308   & 2407 \\\addlinespace[0.1in]
    3C~273 & 1310  & 7485  &       & 1064  & 4254  &       & 363   & 2804 \\\addlinespace[0.1in]
    M~87  & 1017  & 1453  &       & 620   & 620   &       & 190   & 505 \\\bottomrule
     \end{tabular}}
  \label{tab:tbl4}%
\end{table}%

In Table \ref{tab:tbl4}, we compare the dynamic range of each image ($I_\mathrm{p}$/$\sigma_\mathrm{im}$), and the SNR ($I_\mathrm{p}$/$\sigma_\mathrm{th}$), which are derived from the observation parameters of KaVA at 23 and 43 GHz, and VERA at 23 GHz. 
The expected thermal noise $\sigma_\mathrm{th, KaVA}$ is 0.7 \mjybm at 23 GHz, 1.1 \mjybm at 43 GHz, and $\sigma_\mathrm{th, VERA}$ is 1.8 \mjybm at 23 GHz. Although, based on Eq. (1) and a typical parameter for our observation at 23 GHz, we expect $\sigma_{\mathrm{th, KaVA}}$ to be 2.6 times smaller than $\sigma_\mathrm{th, VERA}$, the dynamic range of KaVA images improved to be more than three times that of observations made with only VERA. 

It is clear that KaVA observations sufficiently reduced phase and/or amplitude error, which should affect the image quality, compared with VERA observations. As shown in Table \ref{tab:tbl4}, image dynamic ranges for 4C~39.25 and 3C~273 at 23 and 43 GHz, and for M~87 at 23 GHz are 1000--1800 regardless of the differences in the respective SNR. This is attributed to the phase and/or amplitude errors derived from these brighter sources, which are the dominant factors limiting the dynamic range of the images rather than thermal noise. We find no significant discrepancy between these results and the expected dynamic range.
However, because M 87 at 43 GHz is fainter than the others and its SNR is approximately 600, which is substantially smaller than the expected limit of the dynamic range of the images, it is clear that the image quality of M 87 at 43 GHz is no longer limited by phase and/or amplitude error but rather by thermal noise. Therefore, the image noise level ($\sigma_\mathrm{im}$) is comparable to the expected thermal noise level ($\sigma_\mathrm{th}$) for M~87 at 43 GHz. 
The calculation of the residual phase error ($\Delta\phi$) and the amplitude error ($\epsilon$) of KaVA images according to Eq. (2) gives $\Delta\phi$ of $0.7^{\circ}$ to $1.3^{\circ}$ at 23 GHz and $1.3^{\circ}$ to $2.4^{\circ}$ at 43 GHz, and $\epsilon$ is 1.2\% to 2.2\% at 23 GHz and 2.2\% to 4.2\% at 43 GHz. Comparing with the result of KVN observations reported by \citet{lee14}, both residual phase and amplitude errors of KaVA images are systematically 30\% to 50\% less than those derived from KVN only, except for M~87, (Table \ref{tab:tbl5}). This table also shows the number $M$ of scans for each source and at each frequency.

\begin{table}[htbp]
  \caption{Phase and amplitude calibration error of KaVA/KVN images.}
  \centering
    \scalebox{0.8}{\begin{tabular}{lcccccccc}
    \toprule\toprule
    Source & \multicolumn{3}{c}{23 GHz} &       & \multicolumn{3}{c}{43 GHz} & Array \\\cmidrule{2-4}\cmidrule{6-8}
          & $M$   & $\Delta\phi$ & $\epsilon$ &       & $M$   & $\Delta\phi$ & $\epsilon$ &  \\\addlinespace[0.05in]
          &   (1)    & (2) & (3)   &       &       &      &      &  (4)  \\\midrule
    4C~39.25 & 11    & $<0.7$ & $<1.2$ &       & 13    & $<1.3$ & $<2.3$ & KaVA \\\addlinespace[0.05in]
    3C~273 & 11    & $<0.9$ & $<1.6$ &       & 13    & $<1.3$ & $<2.2$ & KaVA \\\addlinespace[0.05in]
    M~87  & 12    & $<1.3$ & $<2.2$ &       & 16    & $<2.4$ & $<4.2$ & KaVA \\\addlinespace[0.1in]
    NRAO~150 & 12    & $<1.5$ & $<2.7$ &      &   $\cdots$    &   $\cdots$    &   $\cdots$    & KVN \\\addlinespace[0.05in]
    OJ~287 & 12    & $<1.2$ & $<2.0$ &       &   $\cdots$    &   $\cdots$    &   $\cdots$    & KVN \\\addlinespace[0.05in]
    0838+066 & 13    & $<1.6$ & $<2.8$ &       &   $\cdots$    &   $\cdots$    &   $\cdots$    & KVN \\\addlinespace[0.05in]
    0422+022 & 10    & $<1.3$ & $<2.2$ &       &   $\cdots$    &   $\cdots$    &   $\cdots$    & KVN \\\addlinespace[0.05in]
    3C~84 & 148   & $<1.4$ & $<2.4$ &       &   148    & $<1.9$ & $<3.3$ & KVN \\\bottomrule\addlinespace[0.05in]
\multicolumn{8}{@{}l@{}}{\hbox to 0pt{\parbox{105mm}{\footnotesize
\textbf{Notes.~}(1): Number of scan in each observation. (2): Residual phase error of each image in degree. (3) Residual amplitude rms error of each image in percent. (4): Array name. Each parameter of KVN is derived from \citet{lee14}
}\hss}}
    \end{tabular}}%
  \label{tab:tbl5}%
\end{table}

In our KaVA observations, we conducted multiple snapshot observations within an hour angle spread of 6 h (23 GHz) or 8 h (43 GHz) to achieve good $uv$-coverage for each source. Under these conditions, we obtained an image dynamic range exceeding 1000 for brighter sources. However, the dynamic range of images of 4C~39.25 and 3C~273 at 43 GHz is less than that at 23 GHz, which is possibly because of the lack of more than half of the scans from the VERA Iriki and VERA Ishigaki-jima stations because of bad weather conditions (Figures \ref{fig:gain_offset_k}, \ref{fig:gain_offset_q}).

\section{Summary}\label{conclusion}
 To evaluate the imaging capability of a new VLBI array, we conducted KaVA observations of the bright AGNs, 4C~39.25, 3C~273, and M~87 at 23 and 43 GHz. The main results of this work are the following:
\begin{itemize}

\item Our observations clarified that KaVA can achieve the substantially high dynamic range exceeding $\sim1000$ for the sources having the extended structures with even the bandwidth of only 32 MHz and the on-source time of an hour, if we perform the observation as covering the spatial frequency uniformly within an hour angle spread of a total observation time of 6-8 h. The dynamic range derived from KaVA observations is at least three times higher than only VERA observations. This imaging capability of KaVA has a great advantage for the statistical study of AGN jet science such as discussing the jet kinematics effectively by observing many AGN jets as much as possible within the limited time.
 \ \\
 \item By analyzing three sources that have different structures, we found that the scaling factors of antenna gain of all seven stations differ by less than 10\%, with standard deviations of 4\% to 7\% for each source. Therefore, we conclude that no large systematic gain error are found between KVN and VERA, and that the uncertainties in the gain calibration of KaVA is less than 10\%. This is one of the major reasons that we can obtain the high-quality VLBI images by KaVA observations.
 In addition, we conducted two types of amplitude calibrations: an "\textit{a prori}" calibration and a "template" method. Upon comparing the two, we find a tiny difference in the calibration tables (less than 5\%), which is sufficiently smaller than expected calibration error. 
 \ \\
 \item We also generated model-fit images by fitting two-dimensional Gaussian models to the visibility data, and derived spectral index distributions for each source based on the quasi-simultaneous dual frequency observations. Compared with only VERA observations, the number of jet components fit by two-dimensional Gaussian components increased significantly, especially for 3C 273 and M 87. For the spectral index maps, the results derived from KaVA observations are consistent with those published by other groups. 

\end{itemize}

In addition, we mention our prospect below as follows:
\begin{itemize}
\item Because KaVA is verified by this work to offer superior imaging capability, we will consider further AGN jet science that involves KaVA as a large project of an AGN subworking group. One possible project is to monitor several AGNs by using KaVA at an interval of less than one month (typically from biweekly to monthly) at 23 and/or 43 GHz, such as the \textit{GENJI} programme \citep{nagai13}. 
 Currently we call this project the "extended \textit{GENJI}" (e-\textit{GENJI}). In the millimeter wavelength range, we may see emission from the radio core itself by avoiding SSA in the jets. In the future, we also expect to monitor the spectral index distribution of each source by quasi-simultaneous observations at 23 and 43 GHz. 
 Although astrometry with KaVA is currently under commissioning, AGN astrometry with KaVA is also one of the important topic to extend the AGN science. 
\ \\
\item Future 1 Gbps data sets from KaVA regular operation will be correlated by the KJJVC installed at the KJCC located at KASI. In the East Asia region, KaVA is a systematic and promising millimeter VLBI array with time available for common use. The regular operation of KaVA should become the foundation of the East Asia VLBI Network (EAVN).
\end{itemize}


\bigskip
%
We are grateful to all staff members and students at the KVN and VERA who
helped to operate the array and to correlate the data. 
The KVN is a facility operated by the Korea Astronomy and Space Science Institute.
VERA is a facility operated by National Astronomical Observatory of Japan in
collaboration with Japanese universities. 
%

\appendix
\section{Systematic Gain Error}\label{appendix}
\begin{figure*}[htbp]
\centering
\begin{minipage}{0.48\hsize}
\begin{center}
\includegraphics[width=0.95\linewidth]{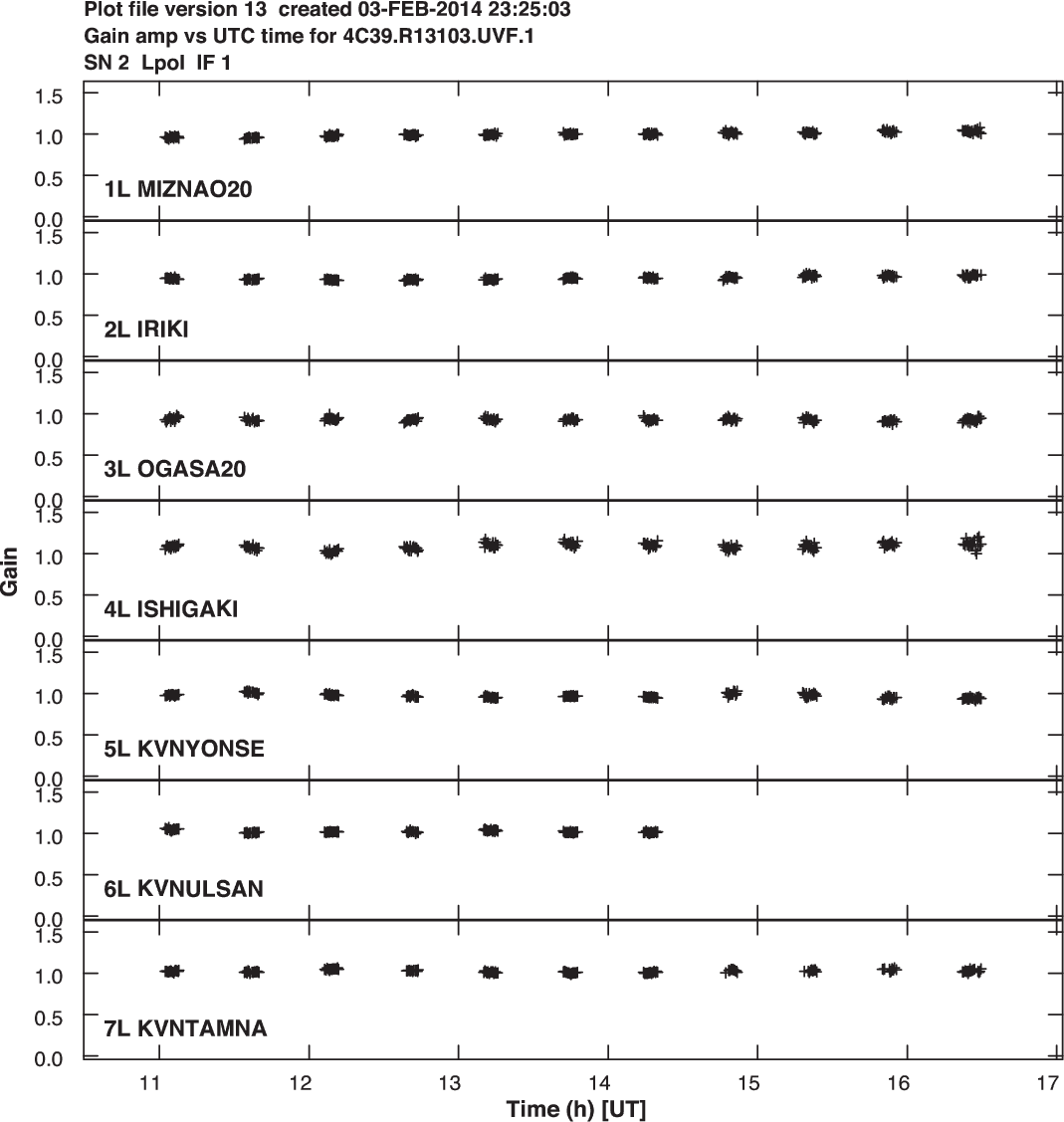}
\end{center}
\caption{Scaling factor for antenna gain derived from self-calibration for 4C 39.25 in the K band.}\label{fig:gain_offset_k}
\end{minipage}
\begin{minipage}{0.48\hsize}
\begin{center}
\includegraphics[width=0.95\linewidth]{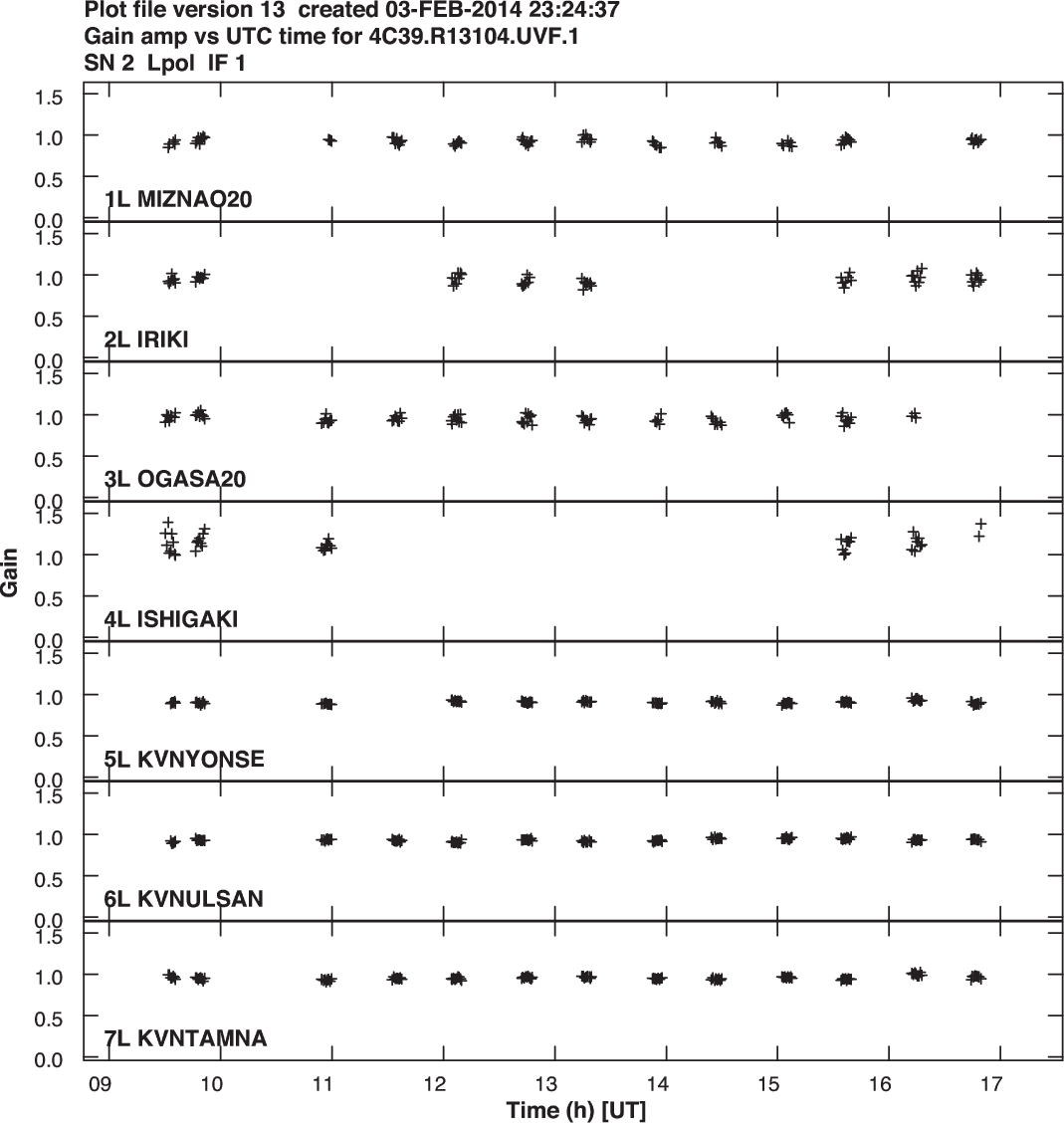}
\end{center}
\caption{Scaling factor for antenna gain derived from self-calibration for 4C 39.25 in the Q band.}\label{fig:gain_offset_q}
\end{minipage}
\end{figure*}

%
To verify whether the systematic antenna gain offset can be detected between KVN and VERA, we compared antenna gain scaling factors derived from an amplitude self-calibration procedure for all sources among all stations at 23 and 43 GHz (Figures \ref{fig:gain_offset_k} and \ref{fig:gain_offset_q}). 
Relative to that with self-calibration, the scaling factor of each visiblity amplitude without self-calibration is calculated by using the AIPS task CALIB. In this process, we set a solution interval to 20 s at 23 GHz and to 30 s at 43 GHz (for M 87 in the 43 GHz band, we set the solution interval to 1 min because its peak intensity is faint, $<0.7\jybm$, compared with other sources)

\begin{table*}[htbp]
  \centering
  \caption{Correction factors of antenna gain derived from self-calibration for each source.}
    \scalebox{0.85}{\begin{tabular}{lcccccccc}\toprule\toprule
    \multicolumn{1}{l}{Source (Band)} & IF    & KVN YNS & KVN ULS & KVN TMN & VERA MIZ & VERA IRK & VERA OGA & VERA ISG \\\addlinespace[0.05in]
    \multicolumn{1}{c}{} &   (1)    &   (2)    &   (3)    &   (4)    &    (5)   &   (6)    &   (7)    & (8) \\\midrule
    \multicolumn{1}{l}{4C~39.25 (K)} & 1     & 0.97 (0.03) & 1.02 (0.02) & 1.03 (0.02) & 1.00 (0.03) & 0.95 (0.02) & 0.93 (0.02) & 1.09 (0.04) \\\addlinespace[0.05in]
    \multicolumn{1}{c}{} & 2     & 0.97 (0.03) & 1.03 (0.02) & 1.03 (0.02) & 0.99 (0.03) & 0.98 (0.01) & 0.93 (0.02) & 1.09 (0.04) \\\addlinespace[0.1in]
    \multicolumn{1}{l}{3~C273 (K)} & 1     & 0.96 (0.03) & 1.01 (0.02) & 1.03 (0.02) & 0.99 (0.02) & 0.95 (0.02) & 0.96 (0.03) & 1.09 (0.04) \\\addlinespace[0.05in]
    \multicolumn{1}{c}{} & 2     & 0.96 (0.03) & 1.01 (0.01) & 1.03 (0.02) & 0.99 (0.02) & 0.98 (0.02) & 0.96 (0.03) & 1.08 (0.03) \\\addlinespace[0.1in]
    \multicolumn{1}{l}{M~87 (K)} & 1     & 0.93 (0.04) & 1.02 (0.04) & 1.00 (0.04) & 1.00 (0.06) & 0.94 (0.04) & 0.95 (0.09) & 1.14 (0.11) \\\addlinespace[0.05in]
    \multicolumn{1}{c}{} & 2     & 0.90 (0.05) & 1.01 (0.04) & 1.00 (0.05) & 0.99 (0.06) & 0.98 (0.04) & 0.96 (0.07) & 1.11 (0.11) \\\midrule
    \multicolumn{1}{l}{4C~39.25 (Q)} & 1     & 0.93 (0.04) & 0.94 (0.02) & 0.97 (0.03) & 0.94 (0.06) & 0.97 (0.08) & 0.97 (0.06) & 1.24 (0.23) \\\addlinespace[0.05in]
    \multicolumn{1}{c}{} & 2     & 0.90 (0.04) & 0.95 (0.02) & 0.96 (0.02) & 0.97 (0.06) & 1.02 (0.10) & 1.01 (0.07) & 1.28 (0.24) \\\addlinespace[0.1in]
    \multicolumn{1}{l}{3~C273 (Q)} & 1     & 0.91 (0.02) & 0.94 (0.01) & 0.97 (0.02) & 0.94 (0.06) & 0.94 (0.06) & 0.95 (0.03) & 1.22 (0.19) \\\addlinespace[0.05in]
    \multicolumn{1}{c}{} & 2     & 0.89 (0.02) & 0.95 (0.01) & 0.96 (0.02) & 0.96 (0.06) & 0.98 (0.06) & 0.97 (0.04) & 1.32 (0.19) \\\addlinespace[0.1in]
    \multicolumn{1}{l}{M~87 (Q)} & 1     & 0.89 (0.07) & 0.92 (0.05) & 0.95 (0.09) & 0.93 (0.09) & 1.00 (0.17) & 0.98 (0.16) & 1.22 (0.43) \\\addlinespace[0.05in]
    \multicolumn{1}{c}{} & 2     & 0.87 (0.05) & 0.93 (0.05) & 0.94 (0.07) & 0.95 (0.09) & 1.04 (0.14) & 0.97 (0.15) & 1.34 (0.34) \\\bottomrule\addlinespace[0.05in]
\multicolumn{9}{@{}l@{}}{\hbox to 0pt{\parbox{185mm}{\footnotesize
\textbf{Notes:~} The value in parentheses is the standard deviation of the correction factor for each antenna. (1) Number of IF bandwidth channels. (2) KVN Yonsei station. (3) KVN Ulsan station. (4) KVN Tamna station. (5) VERA Mizusawa station. (6) VERA Iriki station. (7) VERA Ogasawara station. (8) VERA Ishigaki-jima station.\\
}\hss}}
    \end{tabular}}%
  \label{tab:tbl6}%
\end{table*}%
%

%
%

Based on these comparisons, no significant difference in antenna gain scaling factor for each source was found between both individual arrays. The differences of the averaged scaling factors relative to 1 are less than 10\%, and the standard deviation for each antenna is also within 10\% for all sources (Table \ref{tab:tbl6}).
However, for the VERA Ishigaki-jima station, an apparent systematic offset and a large standard deviation are observed for all sources, especially in the 43 GHz band. However, we attribute this large systematic gain offset to bad weather conditions at the time of observations. Therefore, we conclude that KaVA consists of seven radio telescopes with almost the same antenna specifications and is suitable for imaging observations. This means that KaVA offers significant advantages for VLBI studies, especially for AGN jets.





\end{document}